\documentclass[fleqn,usenatbib,linenumbers]{mnras}

\usepackage{newtxtext,newtxmath}
\usepackage[dvipsum]{xcolor}

\usepackage[T1]{fontenc}

\DeclareRobustCommand{\VAN}[3]{#2}
\let\VANthebibliography\thebibliography
\def\thebibliography{\DeclareRobustCommand{\VAN}[3]{##3}\VANthebibliography}

\usepackage{graphicx}	
\usepackage{amsmath}	
\usepackage{mathtools}
\usepackage{enumitem}
\usepackage{booktabs,caption}
\usepackage[flushleft]{threeparttable}
\usepackage{subcaption}
\usepackage{academicons}

\def\hi{{{\rm H}\,{\sc i}~}}
\def\hin{{{\rm H}\,{\sc i}}}
\def\mgii{{{\rm Mg}\,{\sc ii}~}}
\def\msun{{M$_{\odot}$}}

\defcitealias{Das2020b}{D20}
\defcitealias{Pingel2018}{P18}

\title[Diffuse \hi in the CGM]{Detection of diffuse \hi emission in the circumgalactic medium of NGC\,891 and NGC\,4565 - II}

\author[Das et al.]{
Sanskriti Das \href{https://orcid.org/0000-0002-9069-7061}{\textcolor{green}{\includegraphics[width=9pt]{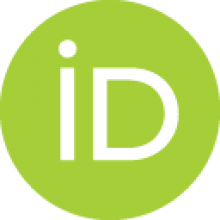}}},$^{1,2}$\thanks{snskriti@stanford.edu}
Mary Rickel \href{https://orcid.org/0009-0007-9943-1183}{\textcolor{green}{\includegraphics[width=9pt]{Orcid-ID.png}}},$^{2}$
Adam Leroy,$^{2,3}$
Nickolas M. Pingel \href{https://orcid.org/0000-0001-9504-7386}{\textcolor{green}{\includegraphics[width=9pt]{Orcid-ID.png}}},$^{4,5}$
D. J. Pisano,$^{6}$
\newauthor George Heald,$^{7}$
Smita Mathur \href{https://orcid.org/0000-0002-4822-3559}{\textcolor{green}{\includegraphics[width=9pt]{Orcid-ID.png}}},$^{2,3}$
Joshua Kingsbury,$^{2}$
and Amy Sardone$^{2}$
\\
$^{1}$Kavli Institute for Particle Astrophysics \& Cosmology, Stanford University, 452 Lomita Mall, Stanford, CA 94305, USA\\
$^{2}$Department of Astronomy, The Ohio State University, 140 West 18th Avenue, Columbus, OH 43210, USA\\
$^{3}$Center for Cosmology and Astroparticle Physics, 191 West Woodruff Avenue, Columbus, OH 43210, USA\\
$^{4}$Research School of Astronomy \& Astrophysics, Australian National University, Canberra, ACT 2611, Australia\\
$^{5}$Department of Astronomy, The University of Wisconsin--Madison, 475 N. Charter Street, Madison, WI 53706, USA \\
$^{6}$Department of Astronomy, University of Cape Town, Rondebosch 7700, Western Cape, South Africa\\
$^{7}$CSIRO Astronomy and Space Science, PO Box 1130, Bentley, WA 6102, Australia
}

\begin{document}
\label{firstpage}
\pagerange{\pageref{firstpage}--\pageref{lastpage}}
\maketitle

\begin{abstract}
We probe the neutral circumgalactic medium (CGM) along the major axes of NGC\,891 and NGC\,4565 in 21-cm emission out to $\gtrsim 100$\,kpc using the Green Bank Telescope (GBT), extending \textcolor{black}{our previous} minor axes observations. We achieve an unprecedented $5\sigma$ sensitivity of $6.1\times 10^{16}$ cm$^{-2}$ per 20\,km\,s$^{-1}$ velocity channel. We detect \hi with diverse spectral shapes, velocity widths, and column densities. We compare our detections to the interferometric maps from the Westerbork Synthesis Radio Telescope (WSRT) obtained as part of the HALOGAS survey. At small impact parameters, $> 31-43\%$ of the emission detected by the GBT cannot be explained by emission seen in the WSRT maps, and it increases to $> 64-73\%$ at large impact parameters.
This implies the presence of diffuse circumgalactic \hin. The mass ratio between \hi in the CGM and \hi in the disk is an order of magnitude larger than previous estimates based on shallow GBT mapping. 
The diffuse \hi along the major axes pointings is corotating with the \hi disk. The velocity along the minor axes pointings is consistent with an inflow and/or fountain in NGC\,891 and an inflow/outflow in NGC\,4565. 
Including the circumgalactic \hin, the depletion time and the accretion rate of NGC\,4565 are sufficient to sustain its star formation. In NGC\,891, most of the required accreting material is still missing.   
\end{abstract}



\section{Introduction} \label{sec:intro}
\noindent The circumgalactic medium (CGM), the multiphase gaseous extended halo surrounding the stellar disk and the interstellar medium (ISM), represents a large reservoir of baryonic material and plays a key role in the evolution of galaxies \citep{Putman2012, Faucher-Giguere2023}. The inflow of 
\hin, the raw fuel of star formation, from the intergalactic medium (IGM) to the ISM is regulated by the CGM through bimodal processes \citep{Keres2005,Voort2011,Nelson2013}. In ``hot mode'' accretion, the accreting gas is shock-heated to the virial temperature,
eventually, a fraction of it cools and fragments into \hi clouds and precipitates to form the \hi disk. It is predominant in massive (virial mass $\rm M_{200}>10^{12} M_\odot$) galaxies and galaxies at lower redshift ($z\lesssim 3$). In lower mass galaxies or galaxies at higher redshift, the gas is predicted to accrete through ``cold mode'', i.e., it is never heated to the virial temperature and it flows through filaments/sheets from the IGM to the ISM bypassing any major thermodynamic interaction with the surrounding hot CGM. 

The depletion time, i.e., the ratio of neutral gas (\hi + H$_2$) mass in the ISM and the star formation rate (SFR) of galaxies is too small to sustain star formation over the cosmic time \citep{Sancisi2008}. The stellar mass density surpasses the neutral gas density since $z\lesssim 2$ \citep{Walter2020}. It implies that there has to be an external supply of gas to keep the ISM in a quasi-equilibrium state. The accretion/inflow rate measured from \hi in and immediately around the disk of local (D$<$25\,Mpc) galaxies is \textcolor{black}{more than} an order-of-magnitude smaller than their star formation rate \citep{Kamphuis2022}. It indicates that the rest of the ``missing'' accreting material could be at larger galactocentric distances, i.e., in the CGM. In addition to the accretion, circumgalactic \hi also traces the tidal interaction, ram-pressure stripping, mergers between galaxies, and galactic outflows \citep[e.g.,][]{Hibbard1996, Wolfe2013, Odekon2016, Veilleux2020}. 

Ly$\alpha$ absorption studies of \hi in the CGM of the Milky Way \textcolor{black}{have shown evidence for a large} amount of atomic gas at N(\hin) $\geq 10^{18}$ cm$^{-2}$ \citep[e.g.,][]{Richter2017}. The circumgalactic \hi in external galaxies has been studied using Ly$\alpha$ absorption lines at $z \lessapprox 0.2-0.85$ \citep[e.g.,][]{Borthakur2016,Peroux2022}. These studies have revealed a strong correlation between the atomic gas mass fraction ($\rm M_{HI}/M_\star$) and the impact-parameter-corrected equivalent width of Ly$\alpha$ absorbers, suggesting a physical connection between the atomic ISM and the CGM. 
However, the Ly$\alpha$ absorption lines are often found saturated but undamped, i.e., 10$^{16}$$<$N(\hin)$<$10$^{18.5}\rm cm^{-2}$, a $\approx$3 \textcolor{black}{orders of magnitude} uncertainty. Absorption studies are limited by the availability of bright background sources like quasars and the covering fraction of the Ly$\alpha$ absorbers. The average of many galaxy-quasar pairs is challenging to interpret as the stellar mass, SFR, and inclination of the galaxy disks, and azimuthal angles of CGM absorbers to the disks are often entangled with each other. Also, the sensitivity to Ly$\alpha$ absorbers at $z\approx 0$ with existing instruments is limited, and it is unknown how the accretion history from $z\approx0.2-0.85$ to $z\approx 0$ evolves, if at all. 21-cm \hi emission studies can constrain the large-scale spatial distribution and kinematics of neutral gas in the CGM of an individual $z\approx 0$ galaxy in an unbiased and complete way. 

Over the past decade, there have been several large interferometric programs to map the \hi emission in the CGM of nearby spiral galaxies at high angular resolution and high point source sensitivity, including Hydrogen Accretion in LOcal GAlaxies Survey \citep[\href{https://www.astron.nl/halogas/data.php}{HALOGAS},][]{Heald2011}, The Local Volume \hi Survey \citep[\href{https://www.atnf.csiro.au/research/LVHIS/}{LVHIS},][]{Koribalski2018}, MeerKAT HI Observations of Nearby Galactic Objects; Observing Southern Emitters \citep[\href{https://mhongoose.astron.nl/sample.html}{MHONGOOSE},][]{Sorgho2019} survey, and Widefield ASKAP L-band Legacy All-sky Blind surveY \citep[\href{https://wallaby-survey.org/}{WALLABY},][]{Kleiner2019}. However, due to the finite size of the dishes, there is a minimum possible spacing (i.e., the diameter of each dish) between neighboring telescopes of an interferometric array. This results in a gap in the \textit{uv}-coverage, which is called the ``short-spacing'' problem. It limits the sensitivity of the interferometers to low surface brightness and large angular scales. For example, the smallest baseline length of the Westerbork Synthesis Radio Telescope (WSRT) is 36\,m, which translates to a maximum recoverable angular scale of $20'$. Because extended \hi structures are more likely to survive than small \hi clouds in the CGM \citep{Armillotta2017,Gronke2022}, the ``diffuse'' circumgalactic \hi with low column density across large angular scales might be a significant reservoir of gas around galaxies that interferometers would miss. Single dishes have full \textit{uv}-coverage, allowing detection of structure at all angular scales. Thus, single-dish observations complement interferometric studies of the CGM. 

\begin{figure*}
\renewcommand{\thefigure}{1a}
    \centering
    \includegraphics[width=0.99\textwidth]{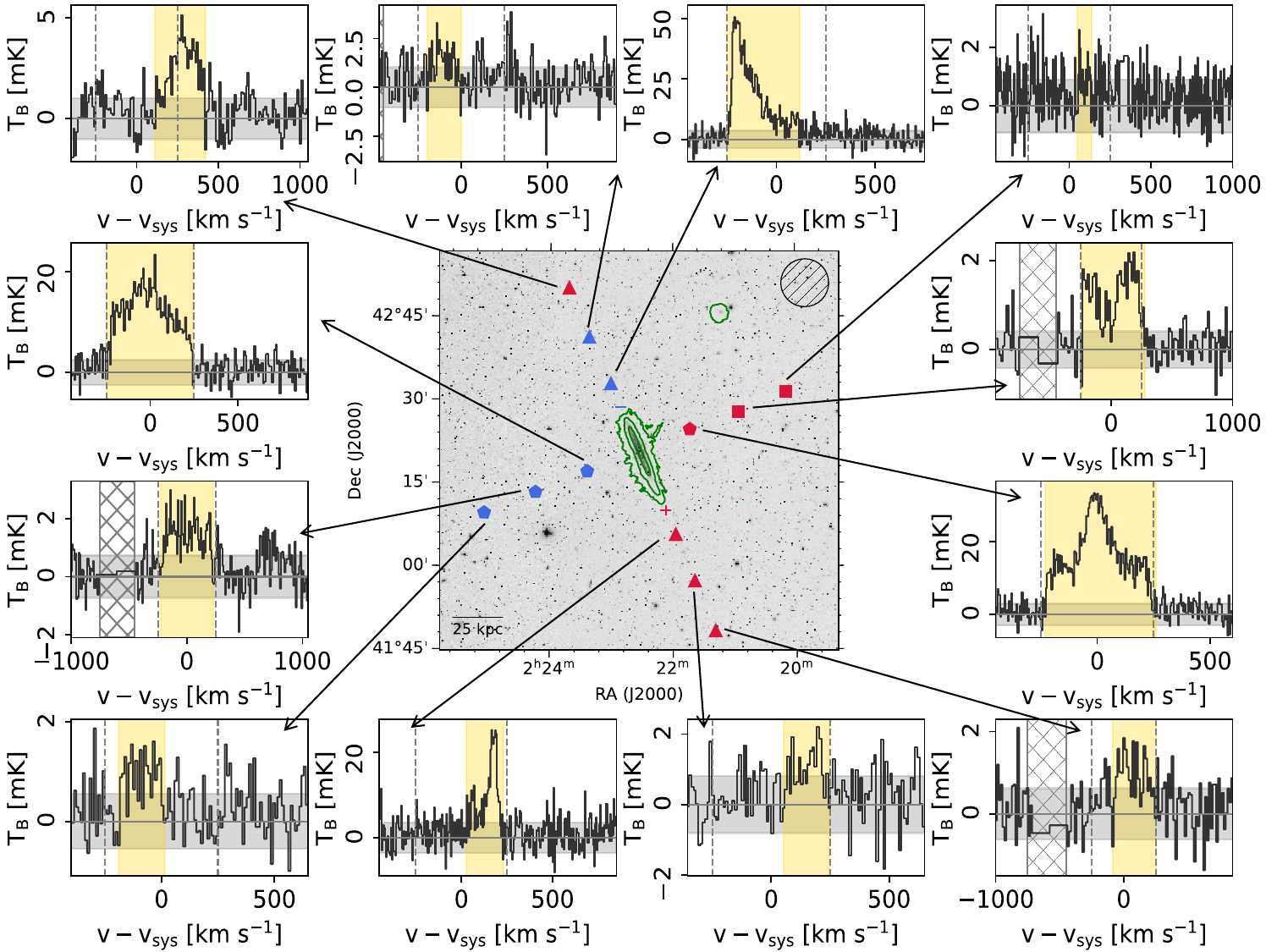}
    \caption{\small The 21-cm emission spectra for the observed GBT pointings in the CGM of NGC\,891, smoothed and re-binned at $\delta$v (see Table \ref{tab:obs}). 
    The green contours on the Digitized Sky Survey (DSS) image show the integrated 21-cm intensity from the HALOGAS survey, with the lowest contour showing an \hi column density of 10$^{19}$ cm$^{-2}$.     
    The GBT pointings observed in 15B-257, 20B-360, and 21B-324 are illustrated with pentagons, triangles, and squares on the DSS image, with red/blue symbols implying the mean line-of-sight velocity of the detected emission being positive/negative (arrows directed from pointings toward respective spectra). The direction of disk rotation has also been marked with `$+/-$' symbols at the edges of the \hi disk.    
    The GBT beam (hatched circle) and the scale are shown at the top and the bottom of the DSS image, respectively.     The companion galaxy UGC\,1807 is visible to the northwest of NGC\,891. 
    In each panel of the spectrum, the maximum rotational velocity of the \hi disk is marked with vertical dashed gray lines, with the systemic velocity of NGC\,891, 528 km s$^{-1}$, subtracted from the actual line-of-sight velocity. 
    The horizontal gray patch denotes the $\pm$RMS value of $\rm T_B$. The horizontal gray line has been drawn at $\rm T_B = 0$ to guide the eye. The vertical hatched region shown in some of the panels is the velocity region of the Milky\,Way that has been masked throughout the data reduction and analysis. 
    We calculate the integrated intensity at each pointing by summing $\rm T_B$ over the velocity range shaded in yellow. At the pointings far from the disk, we detect emission signature at a level of N(\hin) = $1.3 - 13.2 \times10^{17}$ cm$^{-2}$.}
    \label{fig:spectra891}
    \vspace{-0.08 in}
\end{figure*}

The unblocked aperture design of the Green Bank Telescope (GBT), decent angular resolution (FWHM 9.1$'$), its low sidelobes, and high surface brightness sensitivity (T$\rm _{sys} \leqslant$ 20\,K) make it ideal to search for low column density structures over large scales. There have been several single-dish observations with the GBT targeting the CGM of nearby galaxies, as discussed below.

The CGM of NGC\,2403, NGC\,2997 and NGC\,6946 have been mapped with the GBT down to a 5$\sigma$ detection limit of $\sim$10$^{18}$ cm$^{-2}$ over a 20 km s$^{-1}$ channel \citep{deBlok2014,Pisano2014}. Observations of NGC\,2403 revealed a low column density, 8\,kpc extended filament outside the \hi disk. It has recently been found to be part of a 20\,kpc stellar stream that formed due to a recent (2\,Gyr) tidal interaction with a dwarf satellite galaxy \citep{Veronese2023}. The halo of NGC\,2997 did not show any filamentary feature, and the \hi mass, as measured with the GBT, was only 7\% higher than that derived from past interferometric measurements. The \hi observations of NGC\,6946 revealed a filament of N(\hin) = 5 $\times$ 10$^{18}$ cm$^{-2}$ connecting the galaxy with its nearest companions. It shows that the \hi distribution in the CGM varies from galaxy to galaxy.  

As part of the project AMIGA (Absorption Maps In the Gas of Andromeda), \cite{Howk2017} looked for the 21-cm emission with the GBT in the CGM of M\,31 along 48 directions at a 5$\sigma$ sensitivity of 4$\times$10$^{17}$ cm$^{-2}$. Except for the 2 sightlines passing through the Magellanic Stream, they were not able to detect any \hi emission across an impact parameter of 25\,kpc to 340\,kpc. The non-detection indicated that unless M\,31 is an exception, the covering fraction of the circumgalactic \hi might be too low to be detected in the galaxies as nearby as M\,31, where the beam of GBT corresponds to a physical length of 2\,kpc. 

\cite{Pingel2018}, hereafter \citetalias{Pingel2018}, observed the CGM of four galaxies at a distance of 9--18\,Mpc: NGC\,891, NGC\,925, NGC\,4414 and NGC\,4565 with the GBT, and compared it with the interferometric data from the HALOGAS survey at equal spatial and velocity resolution. They did not detect any considerable amount of excess \hi from the GBT data than the WSRT data. It indicated that the column density of diffuse \hin, if present, is lower than their 5$\sigma$ GBT detection limit of 0.9--1.4$\times$10$^{18}$ cm$^{-2}$ over a 20 km s$^{-1}$ line width. 

The aforementioned single-dish surveys integrate for a few minutes per point and cover a large area, usually out to the virial radius of each galaxy. \cite{Das2020b} adopted a complementary approach and integrated for 3--4 hours on-source toward individual pointings along the minor axes of NGC\,891 and NGC\,4565, achieving an unprecedented 5$\sigma$ sensitivity of 8.2$\times$10$^{16}$ cm$^{-2}$ over a 20\,km\,s$^{-1}$ line width. It is more than an order of magnitude deeper observation compared to previous single-dish observations. It led to detections of \hi at each pointing. By comparing the sensitive GBT spectra to the deep HALOGAS WSRT measurements, they found that $\approx50$\% of the emission detected by the GBT cannot be explained by the interferometric maps. This confirmed the presence of low column density diffuse circumgalactic \hi along the minor axes of these galaxies. 

\begin{figure*}
\renewcommand{\thefigure}{1b}
    \centering
    \ContinuedFloat
    \includegraphics[width=0.99\textwidth]{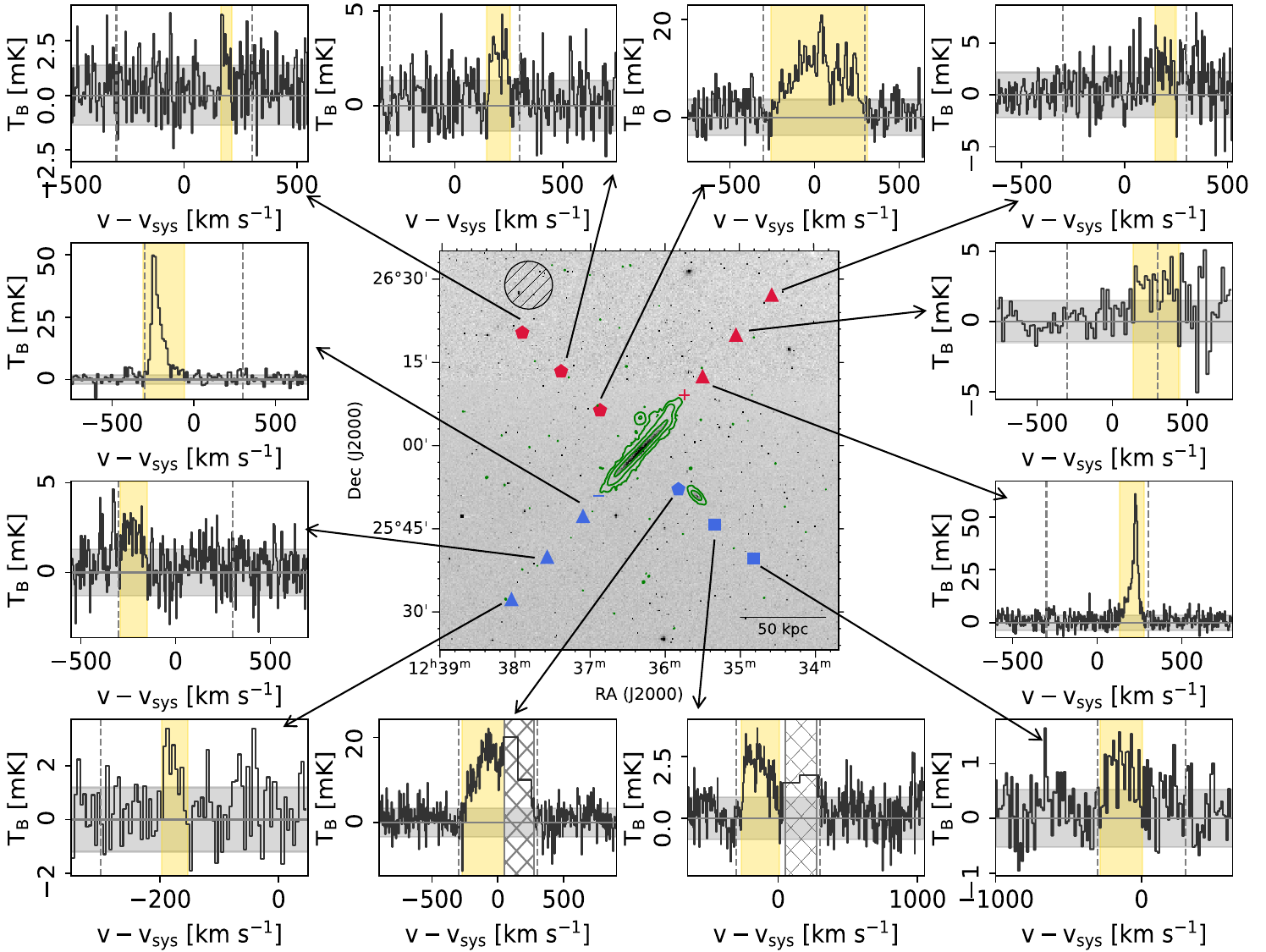}
    \caption{The 21-cm emission spectra for the observed GBT pointings in the CGM of NGC\,4565. The systemic velocity of NGC\,4565, 1230 km s$^{-1}$, has been subtracted from the actual line-of-sight velocity in each spectrum. The companion galaxy NGC\,4562 is visible to the southwest of NGC\,4565. 
    The vertical hatched region shown in some panels of the spectrum is the velocity region of NGC\,4562 that has been masked throughout the data reduction and analysis. See the caption of Fig.\,\ref{fig:spectra891} for more details.}
    \label{fig:spectra4565}
\end{figure*}

In the GBT census of 18 local MHONGOOSE galaxies, the azimuthally averaged cumulative \hi profiles of 15 galaxies do not truncate at large impact parameters \citep{Sardone2021}. This indirectly suggests the presence of diffuse \hi at a large scale. With a 5$\sigma$ \hi column density sensitivity of 10$^{17.9}$ cm$^{-2}$, \cite{Wang2023} mapped the CGM of NGC\,4631 using Five-hundred-meter Aperture Spherical Telescope (FAST) and detected \hi out to 120\,kpc from the galaxy. By comparing it with HALOGAS WSRT measurements, they found that more than one-fourth of the emission detected by the FAST is diffuse. These results are consistent with the findings of \cite{Das2020b}, hereafter \citetalias{Das2020b}.

In this paper, we expand the study of \citetalias{Das2020b} along the major axes of NGC\,891 and NGC\,4565. We achieve a 5$\sigma$ \hi column density sensitivity of 6.1$\times$10$^{16}$ cm$^{-2}$, equivalent with \hi mass sensitivity of $2.1\times 10^5\;\rm M_\odot$, calculated over a 20 km s$^{-1}$ line width. These are among the deepest \hi observations around external galaxies obtained to date in 21-cm \hin. 

We outline our observation and the data reduction in section \ref{sec:obsred}, and the analyses in section \ref{sec:analysis}. We interpret our results in section \ref{sec:discuss}. We summarize our conclusions and comment on future directions in section \ref{sec:conclude}.

\section{Observations and data reduction}\label{sec:obsred}

\begin{table*}
\caption{Details of observations and data reduction}\label{tab:obs}
\begin{threeparttable}
\begin{tabular}{ccccccccccc}
\toprule
Pointing$^{(a)}$ & {RA} & DEC & t$_{\rm eff}^{(b)}$ & T$_{\rm sys}$ & $\sigma_{\rm T_{\rm B,5}}^{(c)}$ & d$_{\perp}$  & $\delta$v$^{(d)}$ & $\Delta$v$_{\rm emit}^{(e)}$ & $\Delta$v$_{\rm base}^{(f)}$ & n$_{\rm fit}^{(g)}$
\\
& (J2000) & (J2000) & (hrs) & (K) & (mK) & (kpc) & (km s$^{-1}$) & (km s$^{-1}$) & (km s$^{-1}$) & 
\\
(1) & (2) & (3) & (4) & (5) & (6) & (7) & (8) & (9) & (10) & (11)
\\
\midrule \vspace{-.15 in} 
\\
\multicolumn{11}{c}{\bf NGC\,891 (2$^h$22$^m$33.6$^s$, 42$^\circ$20$'$58$''$) }
\\
\midrule 
UP1J & 2$^h$23$^m$00.6$^s$ & 42$^\circ$33$'$03$''$ & 0.25 & 19.7 & 2.82 & 35.3 & 5.0 & (-249,117) & (-450,750) & 1 \\ 
UP2J & 2$^h$23$^m$21.9$^s$ & 42$^\circ$41$'$26$''$ & 1.69 & 18.5 & 1.02 & 60.1 & 7.5 & (-197,-1) & (-475,900) & 2\\
UP3J & 2$^h$23$^m$41.6$^s$ & 42$^\circ$50$'$17$''$ & 1.74 & 18.0 & 0.98 & 85.6 & 10.0 & (110,420) & (-400,1050) & 2 \\
UP1N & 2$^h$21$^m$43.5$^s$ & 42$^\circ$24$'$42$''$ & 0.33 & 18.4 & 2.28 & 26.8  & 5.0 & (-230,258) & (-450,600) & 1    \\
UP2N & 2$^h$20$^m$55.7$^s$ & 42$^\circ$27$'$50$''$ & 3.94 & 19.0 & 0.68 & 51.9  & 15.0 & (-244,306) & (-950,1000) & 3 \\
UP3N & 2$^h$20$^m$9.5$^s$ & 42$^\circ$31$'$24$''$  & 4.11 & 19.0 & 0.67 & 76.6   & 5.0 & (50,137) & (-450,1000) & 3\\
DN1J & 2$^h$21$^m$57.3$^s$ & 42$^\circ$05$'$51$''$ & 0.25 & 20.2 & 2.90 & 43.9 & 5.0 & (26,229) & (-450,850) & 2 \\
DN2J & 2$^h$21$^m$38.8$^s$ & 41$^\circ$57$'$29$''$ & 1.82 & 17.5 & 0.93 & 68.1 & 10.0 & (52,246) & (-355,650) & 1 \\
DN3J & 2$^h$21$^m$18.8$^s$ & 41$^\circ$48$'$28$''$ & 1.74 & 16.9 & 0.92 & 94.2 & 15.0 & (-85,248) & (-1000,850) & 3 \\
DN1N & 2$^h$23$^m$23.6$^s$ & 42$^\circ$17$'$07$''$ & 0.33 & 18.4 & 2.29 & 26.7   & 7.5 & (-248,245) & (-450,900) & 3 \\
DN2N & 2$^h$24$^m$14.0$^s$ & 42$^\circ$13$'$21$''$ & 2.53 & 18.0 & 0.81 & 53.6   & 10.0 & (-227,227) & (-1000,1050) & 3\\
DN3N & 2$^h$25$^m$04.0$^s$ & 42$^\circ$09$'$36$''$ & 2.53 & 18.0 & 0.81 & 80.3   & 10.0 & (-189,13) & (-400,650) & 3 \\
\midrule
\multicolumn{11}{c}{\bf NGC\,4565 (12$^h$36$^m$20.9$^s$, 25$^\circ$59$'$23$''$)} \\
\midrule
UP1J  &  12$^h$35$^m$29.8$^s$ & 26$^\circ$12$'$37$''$& 0.25 & 17.6 & 2.52 & 55.3 & 5.0 & (132,278) & (-600,800) & 1 \\
UP2J  & 12$^h$35$^m$02.9$^s$ & 26$^\circ$20$'$07$''$ & 1.78 & 18.3 & 0.98 & 85.6 & 17.5 & (138,445) & (-775,800) & 3\\
UP3J  & 12$^h$34$^m$34.0$^s$ & 26$^\circ$27$'$17$''$ & 1.84 & 18.3 & 0.97 & 115.9 & 5.0 & (147,249) & (-625,525) & 1 \\
UP1N  & 12$^h$36$^m$52.3$^s$ & 26$^\circ$06$'$27$''$ & 0.22 & 17.6 & 2.69 & 31.7   & 7.5 & (-256,311) & (-750,650) & 3   \\
UP2N  & 12$^h$37$^m$23.7$^s$ & 26$^\circ$13$'$27$''$ & 2.81 & 17.2 & 0.73 & 63.0    & 5.0 & (147,259) & (-350,750) & 3\\
UP3N  & 12$^h$37$^m$55.0$^s$ & 26$^\circ$20$'$24$''$ & 2.81 & 17.2 & 0.73 & 94.0    & 5.0 & (162,210) & (-500,550) & 3\\
DN1J & 12$^h$37$^m$05.9$^s$ & 25$^\circ$47$'$26$''$ & 0.25 & 17.5 & 2.51 & 48.9 & 12.5 & (-309,-55) & (-750,700) & 3 \\
DN2J & 12$^h$37$^m$34.5$^s$ & 25$^\circ$40$'$04$''$ & 1.76 & 18.1 & 0.98 & 79.6 & 5.0 & (-293,-148) & (-550,700) & 3\\
DN3J & 12$^h$38$^m$03.0$^s$ & 25$^\circ$32$'$24$''$ & 1.85 & 18.2 & 0.96 & 111.0 & 5.0 & (-196,-153) & (-350,50) & 2 \\
DN1N & 12$^h$35$^m$49.4$^s$ & 25$^\circ$52$'$13$''$ & 0.22 & 17.6 & 2.67 &  31.3    & 5.0 & (-269,50) & (-900,900) & 3\\
DN2N & 12$^h$35$^m$20.5$^s$ & 25$^\circ$45$'$49$''$ & 4.05 & 17.7 & 0.63 &  59.9  &  5.0 & (-264,11) & (-650,1050) & 3\\
DN3N & 12$^h$34$^m$49.1$^s$ & 25$^\circ$39$'$43$''$ & 4.23 & 17.8 & 0.62 &  89.2  & 12.5 & (-285,5) & (-1000,625) & 3 \\
\bottomrule
\end{tabular}
\begin{tablenotes}
\footnotesize
\item {$(a)$ ``UP'' and ``DN'' denote offset to higher and lower declination than the center of the target galaxy. Pointings along the major and minor axes of the target galaxy end with ``J'' and ``N'', respectively. Pointing ``nJ'' is n$\times$9.1$'$ away from the edge of the \hi disk measured in interferometric surveys (see Figure \ref{fig:spectra891}), and pointing ``nN'' is n$\times$9.1$'$ away from the center of the galaxy, where 9.1$'$ is the FWHM size of the GBT beam. The transverse distances of the pointings from the center of the target galaxy are provided in the 7th column, adopting a distance of 9.2\,Mpc/10.8\,Mpc for NGC\,891/NGC\,4565 from \cite{Heald2012}.}
\item {$(b)$ Effective exposure time t$_{\rm eff}$ = t$_{\rm on}\times$t$_{\rm off}$/(t$_{\rm on}$+t$_{\rm off}$). Where t$_{\rm on}$ and t$_{\rm off}$ are the integration times toward the on-source and the off-source pointings, respectively. To maximize t$_{\rm eff}$, we maintain t$_{\rm on}$ and t$_{\rm off}$ to be the same.}
\item {$(c)$} The expected noise at $\delta \rm v = 5$\,km\,s$^{-1}$ based on the radiometer equation.
\item {$(d)$} The velocity resolution at which the spectrum is smoothed.

\item {$(e)$ The velocity range over which the \hi emission is detected.}

\item {$(f)$ The velocity range that we use to estimate the baseline.}

\item {$(g)$ The order of the best-fitted polynomial to describe the baseline.}
\end{tablenotes}
\end{threeparttable}
\end{table*}

We observed the CGM of NGC\,4565 and NGC\,891 in 21-cm (1.42 GHz) emission 
as part of GBT projects 20B-360 (PI: Das) and 21B-324 (PI: Das). These observations add a total of eight new, deep spectra around NGC\,891 and eight new, deep spectra around NGC\,4565 (Figure \ref{fig:spectra891}, triangles and squares). We combine these with observations from GBT project 15B-257 (Figure \ref{fig:spectra891}, pentagons) published in \citetalias{Das2020b}. Including all observations, there are three pointings on each side of each target, with pointings stepping along the major axes (20B-360) and minor axes (15B-257, 21B-324), separated from one another by the FWHM beam size.

To minimize the contamination of the observed spectrum by emission from the disk, we choose the major axes pointings at least 1 FWHM beam size away from the edge of the \hi disk measured in interferometric surveys (Figure \ref{fig:spectra891}, contours), and minor axes pointings at least 1 FWHM beam-size away from the center of the galaxy. The details of the observations, including the sky direction of the pointings, the effective integration time at each pointing, $\rm t_{eff}$, and the system temperature, T$_{\rm sys}$ are provided in Table \ref{tab:obs}. 

\begin{figure*}
    \centering
    \includegraphics[width=0.48\textwidth]{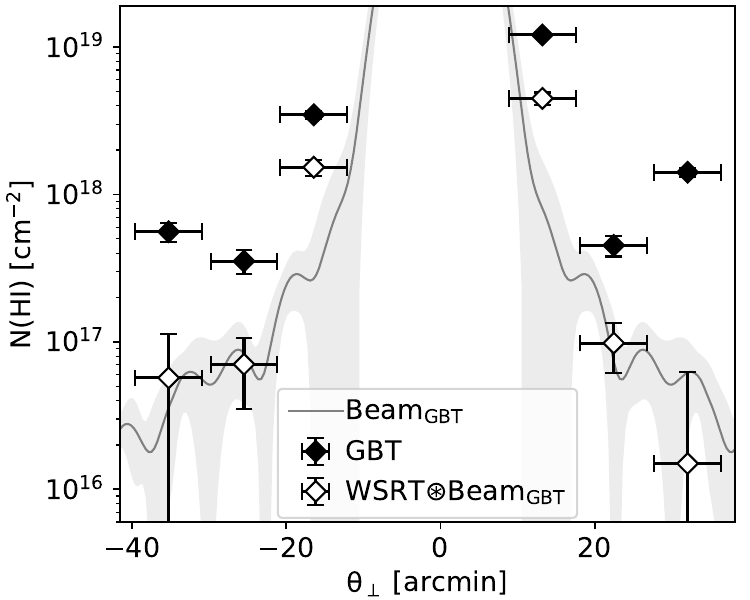}
    \includegraphics[width=0.48\textwidth]{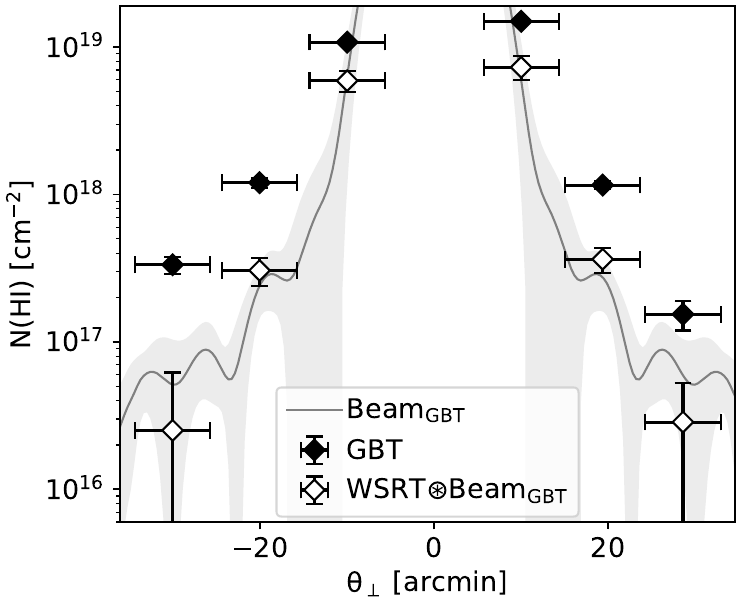}
    \includegraphics[width=0.48\textwidth]{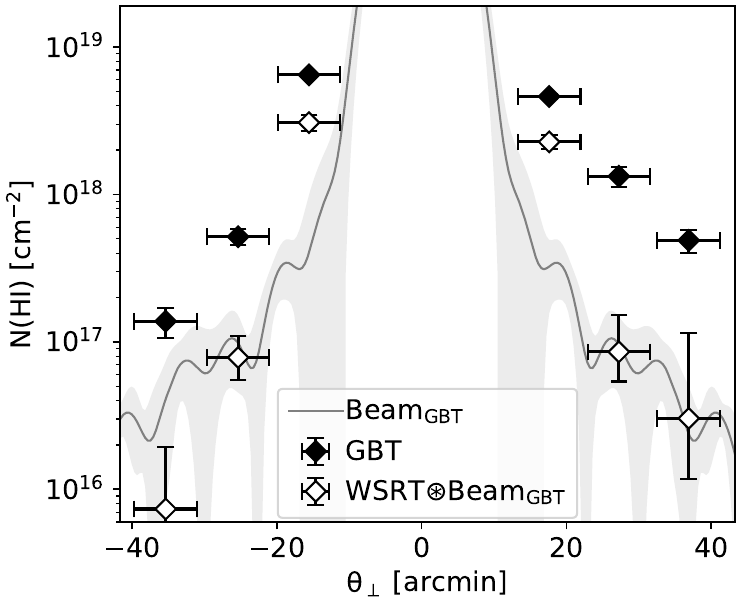}
    \includegraphics[width=0.48\textwidth]{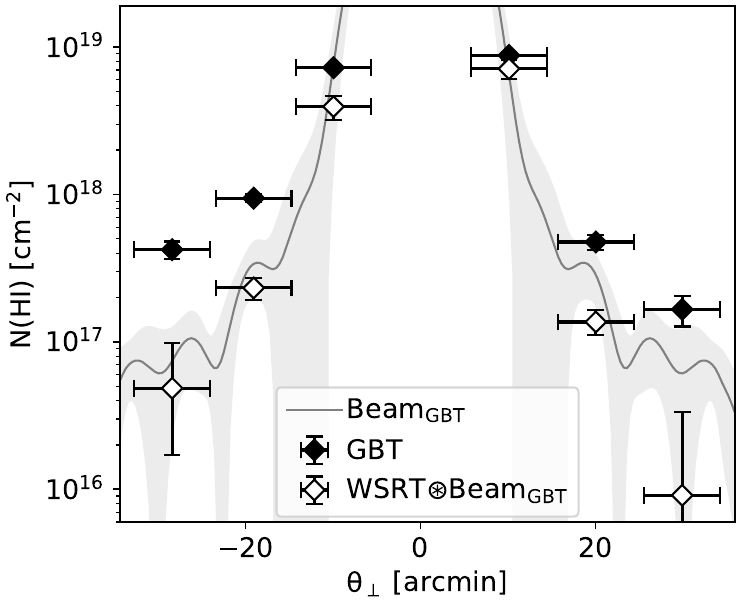}
    \caption{The radial N(\hin) profile of NGC\,891 (top) and NGC\,4565 (bottom) along their major axes (left) and minor axes (right). The pointings at higher/lower declination than the galaxy disk are plotted at positive/negative angular separation from the galaxy center, respectively. The error bars along the y-axis include systematic uncertainties and statistical uncertainties. The error bars along the x-axis denote the GBT beam size. 
    In our GBT data, systematic uncertainties come from the difference in $\rm T_{cal}$ across different sessions. 
    In the WSRT data, systematic uncertainties come from the difference in masking threshold (S/N$>$4--5--6), and circularization of the GBT beam. The shaded region around the beam model is 1$\sigma$ uncertainty in the beam response due to averaging the beam map to one dimension. The positive offset between our GBT data and the convolved WSRT data indicates the presence of diffuse circumgalactic \hin.}
    \label{fig:NH}
\end{figure*}

We observed by position switching, using the L-band receiver with the Versatile GBT Astronomical Spectrometer (VEGAS; bandwidth = 23.5 MHz) as the backend. The off-source pointings are chosen to have the same declination as the target pointings and are placed $\approx1.5^\circ$ away from the disk. This places the off-source pointing beyond the virial radii of the target galaxies. We also make sure that there is no known \hi source within 1 FWHM beam size of the off-source pointings in the velocity range of concern. We used the quasars 3C\,48 (in 15B-257, 20B-360) and 3C\,123 (in 21B-324) as the primary flux calibrator for NGC\,891 and 3C\,286 for NGC\,4565. The \hi disk of each galaxy was observed before the observations of the CGM to verify the setup.

In project 20B-360, for the calibrator source of 3C\,48 (3C\,286), the average effective temperature of the noise diode, T$_{\rm cal}$ is 1.60$\pm$0.01\,K (1.45$\pm$0.09\,K) for the XX polarization and 1.58$\pm$0.01\,K (1.66$\pm$0.09\,K) for the YY polarization. When the polarizations are averaged to derive the Stokes\,I component, this translates to $0.6\% (0.7\%)$ systematic uncertainty due to uncertain flux calibration. The systematic uncertainties in projects 21B-324 and 15B-257 are $1.4\%(1.4\%)$ and $0.2\%(1.0\%)$, respectively. 

We use an improved version of the routine developed in GBTIDL by \citetalias{Das2020b} to reduce the position-switched data. 
We obtain the aperture efficiency and the atmospheric opacity at the zenith, which are used to derive T$_{\rm cal}$, based on the elevation during each integration and at each frequency. We use that to construct an off-source spectrum from each off-source scan, then subtract this from the on-source spectrum corresponding to the adjacent on-source scan.
The calibrated, off-subtracted spectrum of each scan in a session is then weighted by the combined integration time and system temperature of the on-off pair, t$_{\rm eff}$/T$_{\rm sys}^2$, to calculate a combined spectrum.
As part of the processing, we also identify channels visibly affected by RFI (radio frequency interference) and replace these with values linearly interpolated from nearby RFI-free parts of the spectrum.

We construct a spectrum at each pointing in each session for each polarization in units of brightness temperature, T$_{\rm B}$. We shift the velocity axis of each spectrum so that v $ = 0$ km s$^{-1}$ corresponds to the systemic velocity of the galaxy, v$_{\rm sys}$. This is 528 km s$^{-1}$ for NGC\,891 and 1230 km s$^{-1}$ for NGC\,4565 \citep{Rupen1991}. We accumulate the spectra from individual sessions at the native velocity resolution of 1.25 km s$^{-1}$.

For each pointing, we combine the spectra for each polarization from all sessions, again weighted by the total t$_{\rm eff}$/T$_{\rm sys}^2$ for each single-session spectrum. Thus, we obtain the mean spectrum for each polarization, in units of brightness temperature. Finally, we average the two polarizations, which increases the final S/N by a factor of $\sqrt{2}$. 

After constructing the final spectrum, we apply an iterative signal finding and baseline fitting technique, which we describe in Appendix \ref{sec:baseline}. The algorithm tests a variety of channel widths and baseline orders. It identifies regions of candidate emission, fits regions away from candidate emission, and subtracts the lowest order robustly-fitted polynomial baseline. Table \ref{tab:obs} reports the optimum channel width $\delta v$, velocity range used for the baseline fit $\rm \Delta v_{base}$, the velocity range of identified emission $\rm v_{emit}$, and the order of the polynomial fit $n_{\rm fit}$. Because this differs somewhat from the approach used by \citet{Das2020b}, we also re-reduce, re-stack, and then fit new baselines to the spectra from that paper (project 15B-257). This gives us a single homogeneously reduced data set.

As a final check, we compare the RMS noise measured from signal-free regions of the baseline-subtracted spectrum with the noise expected based on the radiometer equation using t$_{\rm eff}$ and T$_{\rm sys}$ (Table \ref{tab:obs}). The measured RMS noise is on average a factor of $1.18_{-0.12}^{+0.08}$ and $1.36_{-0.10}^{+0.60}$ larger than the expected noise for NGC\,891 and NGC\,4565. This could be due to residual imperfections in the baseline subtraction that require higher-order polynomial or more sophisticated baseline fitting techniques. Below we use the measured RMS noise rather than the radiometer equation-based estimates, so we expect our S/N estimation to be conservative and capture this additional uncertainty. 

\section{Analysis}\label{sec:analysis}
In the final spectra, we detect \hi emission in most of the pointings (Figure \ref{fig:spectra891} and \ref{fig:spectra4565}, yellow shaded area). 
We integrate each spectrum over the range of $\Delta$v$_{\rm emit}$ to obtain the intensity, $\rm \int T_B dv$ in units of K km s$^{-1}$. We calculate the statistical uncertainty in the integrated intensity using $\sigma_{\rm T_B}$, $\delta$v and $\Delta$v$_{\rm emit}$. We add the systematic uncertainty in flux calibration (see \S\ref{sec:obsred}) to the statistical uncertainty in quadrature. Except for the pointings immediately next to the \hi disk, our detections at all other pointings lie below the $5\sigma$ sensitivity limit of the GBT-HALOGAS survey \citepalias{Pingel2018}. 

We convert the integrated intensity at each pointing to column density, N(\hin)$\rm_{GBT}$, assuming that the emission is optically thin and the number density of the gas is well above the critical density of \hi for collisions with electrons \citep[7$\times$10$^{-6}$ cm$^{-3}$ at 10$^4$ K;][]{Draine2011}. 
We do not apply any correction for the inclination angle in computing the column density, because we do not know the geometry and morphology of the detected emission. We show the derived N(\hin) as a function of galactocentric radius in Figure \ref{fig:NH}. 

We calculate the intensity-weighted mean velocity over the range of $\Delta$v$_{\rm emit}$: $\rm {\bar{v}}_{GBT} = \frac{\rm \int T_Bv dv}{\int \rm T_B dv}$. We also calculate the \hi mass within each beam from the integrated intensities, adopting a gain of 1.86 K Jy$^{-1}$. We quote the N(\hin)$\rm_{GBT}$, the associated \hi masses, and $\rm\bar{v}_{GBT}$ in Table \ref{tab:results}.  

\subsection{Comparison with interferometric data}
Our GBT spectra include emission both from high column density \hi clouds and any diffuse, extended CGM structures. They may also include stray light from the disk of the galaxy, reflecting that even the relatively clean GBT beam can still pick up emission from the bright inner region of the galaxy. To help disentangle these components, we compare our GBT measurements to sensitive interferometric maps from the HALOGAS survey using WSRT \citep{Heald2011}. These interferometric data are sensitive to compact, high column density structures and they capture the main disk of the galaxy well, but they lack sensitivity to extended, low surface brightness structures. 

To carry out a rigorous comparison, we convolve the masked interferometric data cubes with the GBT beam so that the interferometric data are at the angular resolution of the GBT. From the convolved interferometric maps, we can estimate the contribution of compact, relatively high column density structures to our observed GBT spectra, i.e., any extraplanar \hi emission, and pickup of the galaxy disk from the sidelobes of the GBT beam. The excess in the GBT spectra compared to the spectra from convolved WSRT data cubes will correspond to the extended low column density structures missed by the interferometer.

\begin{table*}
\caption{Details of measurements}\label{tab:results}
\begin{threeparttable}
\begin{tabular}{cccccccc}
\toprule
Pointing & N(\hin)$_{\rm GBT}^{(a)}$ & $\bar {\rm v}_{\rm GBT}$ & M(\hin)$_{\rm GBT}$ & N(\hin)$_{\rm WSRT}^{(b)}$ & N(\hin)$_{\rm CGM}^{(c)}$  & $\bar {\rm v}_{\rm CGM}$ & M(\hin)$_{\rm CGM}$ 
\\
& (cm$^{-2}$) &  (km\,s$^{-1}$) & (M$_\odot$) & (cm$^{-2}$) & (cm$^{-2}$) & (km\,s$^{-1}$) & (M$_\odot$)  
\\
(1) & (2) & (3) & (4) & (5) & (6) & (7) & (8)
\\
\midrule \vspace{-.15 in} 
\\
\multicolumn{8}{c}{\bf NGC\,891}\\
\multicolumn{8}{c}{M(\hin)$_{\rm disk}^{(d)}$  =$4.1\times10^9$ M$_\odot$, SFR$^{(d)}$ = 2.2 M$_\odot$yr$^{-1}$, sSFR$^{(e)}$ = $4.0\pm 1.0\times10^{-11}$ yr$^{-1}$, $\Sigma_{\rm SFR}^{(f)} =3.6\times10^{-3}$ M$_\odot$ yr$^{-1}$ kpc$^{-2}$} 
\\
\midrule
UP1J & 19.082$\pm$0.010$\pm$0.003 & -123$\pm$5 & 4.1$\pm$0.1$\times10^7$ & 18.652$\pm$0.005$\pm$0.042 & 18.88$\pm$0.03 & -84$\pm$18 & 2.6$\pm$0.2$\times10^7$ \\
UP2J & 17.655$\pm$0.069$\pm$0.003 & -104$\pm$35 & 1.5$\pm$0.2$\times10^6$ & 16.992$\pm$0.153$\pm$0.053 & 17.55$\pm$0.10 & -91$\pm$30 & 1.2$\pm$0.3$\times10^6$ \\
UP3J & 18.151$\pm$0.032$\pm$0.003 & 282$\pm$1 & 4.8$\pm$0.3$\times10^6$ & 16.175$\pm$1.375$\pm$0.038 & 18.15$\pm$0.04 & 265$\pm$13 & 4.7$\pm$0.4$\times10^6$ \\
UP1N & 19.175$\pm$0.008$\pm$0.001 & 6$\pm$2 & 5.0$\pm$0.1$\times10^7$ & 18.863$\pm$0.003$\pm$0.081 & 18.89$\pm$0.08 & 23$\pm$2 & 2.6$\pm$0.5$\times10^7$ \\
UP2N & 18.064$\pm$0.026$\pm$0.005 & 28$\pm$8 & 3.9$\pm$0.2$\times10^6$ & 17.561$\pm$0.069$\pm$0.048 & 17.90$\pm$0.05 & 32$\pm$5 & 2.7$\pm$0.3$\times10^6$ \\
UP3N & 17.188$\pm$0.097$\pm$0.005 & 90$\pm$1 & 5.2$\pm$1.2$\times10^5$ & 16.456$\pm$0.356$\pm$0.092 & 17.10$\pm$0.15 & 82$\pm$10 & 4.2$\pm$1.4$\times10^5$ \\
DN1J & 18.541$\pm$0.026$\pm$0.003 & 144$\pm$1 & 1.2$\pm$0.1$\times10^7$ & 18.185$\pm$0.010$\pm$0.053 & 18.29$\pm$0.06 & 111$\pm$2 & 6.5$\pm$1.0$\times10^6$ \\
DN2J & 17.547$\pm$0.080$\pm$0.003 & 154$\pm$1 & 1.2$\pm$0.2$\times10^6$ & 16.848$\pm$0.213$\pm$0.053 & 17.45$\pm$0.11 & 140$\pm$18 & 9.5$\pm$2.5$\times10^5$ \\
DN3J & 17.747$\pm$0.062$\pm$0.003 & 72$\pm$8 & 1.9$\pm$0.3$\times10^6$ & 16.756$\pm$0.429$\pm$0.033 & 17.70$\pm$0.08 & 63$\pm$1 & 1.7$\pm$0.3$\times10^6$ \\
DN1N & 19.034$\pm$0.011$\pm$0.001 & -12$\pm$4 & 3.6$\pm$0.1$\times10^7$ & 18.772$\pm$0.004$\pm$0.070 & 18.69$\pm$0.09 & -44$\pm$13 & 1.6$\pm$0.3$\times10^7$ \\
DN2N & 18.080$\pm$0.033$\pm$0.001 & -2$\pm$10 & 4.1$\pm$0.3$\times10^6$ & 17.484$\pm$0.075$\pm$0.054 & 17.95$\pm$0.05 & -5$\pm$7 & 3.0$\pm$0.4$\times10^6$ \\
DN3N & 17.523$\pm$0.059$\pm$0.001 & -86$\pm$26 & 1.1$\pm$0.2$\times10^6$ & 16.399$\pm$0.631$\pm$0.103 & 17.49$\pm$0.08 & -79$\pm$25 & 1.0$\pm$0.2$\times10^6$ \\
\midrule
\multicolumn{8}{c}{\bf NGC\,4565}\\
\multicolumn{8}{c}{M(\hin)$_{\rm disk}^{(d)}$=$7.3\times10^9$ M$_\odot$, SFR$^{(d)}$ = 0.67 M$_\odot$yr$^{-1}$, sSFR$^{(e)}$ = $8.4\pm 2.1\times10^{-12}$ yr$^{-1}$, $\Sigma_{\rm SFR}^{(f)} =6.9\times10^{-4}$ M$_\odot$ yr$^{-1}$ kpc$^{-2}$} \\
\midrule
UP1J & 18.665$\pm$0.016$\pm$0.003 & 215$\pm$0 & 2.1$\pm$0.1$\times10^7$ & 18.357$\pm$0.004$\pm$0.048 & 18.37$\pm$0.06 & 202$\pm$3 & 1.1$\pm$0.1$\times10^7$ \\
UP2J & 18.124$\pm$0.065$\pm$0.003 & 283$\pm$3 & 6.2$\pm$0.9$\times10^6$ & 16.933$\pm$0.155$\pm$0.116 & 18.09$\pm$0.07 & 273$\pm$40 & 5.8$\pm$1.0$\times10^6$ \\
UP3J & 17.689$\pm$0.080$\pm$0.003 & 191$\pm$1 & 2.3$\pm$0.4$\times10^6$ & 16.481$\pm$0.263$\pm$0.207 & 17.66$\pm$0.10 & 180$\pm$30 & 2.1$\pm$0.5$\times10^6$ \\
UP1N & 18.944$\pm$0.022$\pm$0.004 & 32$\pm$6 & 4.1$\pm$0.2$\times10^7$ & 18.853$\pm$0.003$\pm$0.065 & 18.22$\pm$0.30 & 125$\pm$88 & 7.7$\pm$5.4$\times10^6$ \\
UP2N & 17.678$\pm$0.052$\pm$0.004 & 206$\pm$0 & 2.2$\pm$0.3$\times10^6$ & 17.135$\pm$0.059$\pm$0.057 & 17.53$\pm$0.08 & 195$\pm$20 & 1.6$\pm$0.3$\times10^6$ \\
UP3N & 17.221$\pm$0.102$\pm$0.004 & 181$\pm$1 & 7.7$\pm$1.8$\times10^5$ & 15.959$\pm$0.581$\pm$0.320 & 17.20$\pm$0.12 & 172$\pm$31 & 7.3$\pm$2.0$\times10^5$ \\
DN1J & 18.812$\pm$0.014$\pm$0.003 & -213$\pm$12 & 3.0$\pm$0.1$\times10^7$ & 18.488$\pm$0.004$\pm$0.051 & 18.53$\pm$0.05 & -178$\pm$58 & 1.6$\pm$0.2$\times10^7$ \\
DN2J & 17.714$\pm$0.053$\pm$0.003 & -224$\pm$55 & 2.4$\pm$0.3$\times10^6$ & 16.894$\pm$0.116$\pm$0.079 & 17.64$\pm$0.07 & -210$\pm$43 & 2.0$\pm$0.3$\times10^6$ \\
DN3J & 17.140$\pm$0.102$\pm$0.003 & -179$\pm$83 & 6.4$\pm$1.5$\times10^5$ & 15.866$\pm$0.693$\pm$0.067 & 17.12$\pm$0.11 & -169$\pm$54 & 6.1$\pm$1.6$\times10^5$ \\
DN1N & 18.860$\pm$0.015$\pm$0.004 & -85$\pm$7 & 3.4$\pm$0.1$\times10^7$ & 18.596$\pm$0.003$\pm$0.080 & 18.52$\pm$0.10 & -45$\pm$39 & 1.5$\pm$0.4$\times10^7$ \\
DN2N & 17.975$\pm$0.026$\pm$0.006 & -141$\pm$17 & 4.4$\pm$0.3$\times10^6$ & 17.366$\pm$0.054$\pm$0.050 & 17.85$\pm$0.04 & -128$\pm$21 & 3.3$\pm$0.3$\times10^6$ \\
DN3N & 17.626$\pm$0.059$\pm$0.006 & -138$\pm$41 & 2.0$\pm$0.3$\times10^6$ & 16.686$\pm$0.266$\pm$0.178 & 17.57$\pm$0.08 & -127$\pm$34 & 1.7$\pm$0.3$\times10^6$ \\
\bottomrule
\end{tabular}
\begin{tablenotes}
\footnotesize
\item {$(a)$ The statistical uncertainty is followed by the multiplicative systematic uncertainty due to the variation in flux calibration from session to session.}

\item {$(b)$ The column density obtained from the masked WSRT data convolved with the GBT beam map, estimated over the velocity range where emission is detected in the GBT. Here, the statistical uncertainty is followed by the systematic uncertainty. The latter includes multiplicative systematic uncertainty due to averaging the azimuthally asymmetric GBT beam map to 1-D and additive systematic uncertainty in the masking threshold, added in quadrature.}

\item {$(c)$ The excess \hi detected by the GBT compared to that by the WSRT, i.e., the diffuse N(\hin) in the CGM. The error includes (in quadrature) the statistical uncertainties and systematic uncertainties in N(\hin) measured by the GBT and WSRT.}

\item {$(d)$ Adapted from \cite{Heald2012}}

\item {$(e)$ Stellar mass adapted from \cite{Monachesi2016}}

\item {$(f)$ Obtained from \cite{Pingel2018}}  
\end{tablenotes}
\end{threeparttable}
\end{table*}

Same as \citetalias{Das2020b}, we follow the method of 1) global noise estimation and 2) primary-beam correction of the low-resolution\footnote{$35.\!''5 \times32.\!''2$ for NGC\,891 and $43.\!''7 \times33.\!''7$ NGC\,4565 } WSRT data cubes, 3) masking of the primary-beam corrected WSRT data cubes, 4) circularization of the GBT beam model, 5) convolution of the masked WSRT data cube with the circularized GBT beam, and 6) estimation of the statistical uncertainty in the convolved WSRT spectra. 
The WSRT data cubes masked with S/N$\geqslant5$ are used to calculate the best estimate of the spectra. The additive systematic uncertainty comes from the convolution of the WSRT data cubes that are masked with thresholds of S/N$\geqslant4$ and S/N$\geqslant6$. In Appendix\,\ref{sec:threshold} we discuss why S/N$<4$ is not a good mask. 
The circularized beam map accounts for the continuous change in the orientation of the GBT beam throughout the observation, but it averages out the details of the shape of the GBT beam. To account for the azimuthal asymmetry in the GBT beam, we incorporate a multiplicative systematic uncertainty that depends on the sky position of the pointing and thus is different for every pointing. 

By integrating the masked and convolved WSRT spectrum over $\rm \Delta v_{emit}$ at each pointing, we obtain the \hi column density from  WSRT, N(\hin)$\rm_{WSRT}$. We compare N(\hin)$\rm_{GBT}$ to N(\hin)$\rm_{WSRT}$ in Figure\,\ref{fig:NH}. 
We find two sets of results:

\textbf{1) Outer halo (pointings UP/DN 2 and 3):} There is significant excess in our GBT measurements compared to the WSRT estimates across most of the velocity channels of $\Delta$v$_{\rm emit}$ (Figure\,\ref{fig:diffspec}). The excess emission, N(\hin)$\rm_{GBT}$--N(\hin)$\rm_{WSRT}$, is 75$_{-11}^{+9}$\% and 81$_{-8}^{+5}$\% of the GBT emission, N(\hin)$\rm_{GBT}$, of NGC\,891 and NGC\,4565, respectively. 

Because the column densities at these pointings are way below the sensitivity limit of the WSRT data cubes ($\rm \approx 10^{19} cm^{-2}$), the masked WSRT data cube is unlikely to have any high S/N emission within the main lobe of the GBT beam of these pointings. That means the masked and convolved WSRT spectra at these pointings represent the stray light from the \hi disk and the extraplanar region. Therefore, the offset between the GBT and the WSRT estimates at these pointings is due to the diffuse \hi emission from the CGM. From the diverse shape, velocity width, and amplitude of the emissions, we cannot determine whether it consists of extended \hi structure and/or discrete \hi clouds/clumps. It would require high-sensitivity interferometric measurements to distinguish between these cases. 

\textbf{2) Inner halo (pointings UP/DN 1):} The emission detected in GBT at some of the velocity channels is explained by the WSRT. Thus the excess in GBT than WSRT is narrower than $\Delta \rm v_{emit}$ (Figure\,\ref{fig:diffspec}). The excess emission accounts for 54$\pm$11(42$_{-11}^{+10}$)\% of the GBT emission of NGC\,891 (NGC\,4565).  In addition to the stray light from the \hi disk of the target galaxy and its companions (e.g., NGC\,4562), the masked and convolved WSRT spectra at these pointings represent the small-scale \hi in the extraplanar region. Thus, the offset between our GBT and WSRT columns at these pointings represents the diffuse and large-scale \hi emission from the CGM. We test it more rigorously by redefining noise and masking through intermediate convolution in Appendix\,\ref{sec:intermediate_convol} and confirm this inference.

\subsection{Properties of the CGM}\label{sec:propcgm}
We obtain the \hi column density in the CGM, N(\hin)$\rm _{CGM}$, by subtracting the \hi column density of the masked and convolved WSRT spectrum from that of our GBT spectrum at each pointing:
\begin{equation}\label{eq:nhcgm}
  {\rm      N(HI)_{CGM}}(r) = {\rm N(HI)_1}(r) - {\rm N(HI)_2}(r) \rm \frac{N(HI)_{disk,1}}{N(HI)_{disk,2}}
\end{equation}
The subscripts 1 and 2 correspond to the GBT and the WSRT, respectively. The column density of the \hi disk, N(\hin)$\rm _{disk}$, is not the same across all observing sessions of the GBT due to systematic errors. Also, they are not the same as the N(\hin)$\rm _{disk}$ estimated from the masked and convolved WSRT spectrum. Therefore, we apply their ratio as a correcting factor in the latter term in equation\,\ref{eq:nhcgm}. 

We fit the radial profile of N(\hin)$\rm _{CGM}$ as an exponentially decreasing profile:
\begin{equation}\label{eq:density}
   {\rm  N(HI)_{CGM}}(r) = {\rm N(HI)_o exp}(-r/r_s)
\end{equation} 
The two parameters N(\hin)$\rm _o$ and $r_s$ are the central column density of \hi and the scale radius. We also fit different subsets of the data to test the azimuthal symmetry and axial symmetry around the \hi disk.

Using the best-fitted N(\hin)$\rm _{CGM}$ profile from equation\,\ref{eq:density}, we calculate the \hi mass in the CGM \textcolor{black}{by integrating the surface mass density, $m_p {\rm N(HI)_{CGM}}(r)$, over the area of concern:}
\begin{subequations}\label{eq:mass}
\begin{equation}
  {\rm  M(HI)_{CGM}}(r) =  \int m_p {\rm N(HI)_{CGM}}(r) \rm dA  
\end{equation}
\textcolor{black}{To calculate the mass within the main lobe of the GBT beam, we use the area} 
\begin{equation}
{\rm A}(r)_{\rm Beam} = 0.25\pi(\Delta r)^2,
\end{equation}   
\textcolor{black}{with $\Delta r$ being the FWHM of the GBT beam. To calculate the mass in each quadrant}, i.e., within $\pm\pi/4$ of the major and minor axes, we use the area  
\begin{equation}
{\rm A}(r) = 0.5\pi r \Delta r
\end{equation} 
\end{subequations}

We obtain the average line-of-sight velocity of the \hi emission in the CGM, $\rm \bar{v}_{CGM}$, using equation \ref{eq:vavg}.
\begin{equation}\label{eq:vavg}
   \rm \bar{v}_{CGM} = \frac{\int T_{B,1}v_{1} dv_{1} - \int T_{B,2}\circledast Beam_{1}v_{2} dv_{2}}{\int T_{B,1} dv_{1} - \int T_{B,2}\circledast Beam_{1} dv_{2}}
\end{equation}
Here, brightness temperature is integrated over the velocity range of $\rm \Delta v_{emit}$, and subscript 1 and 2 corresponds to the GBT and the WSRT, respectively. 

We calculate the accretion rate using equation \ref{eq:accretion}.  The assumed location of the circumgalactic \hi is illustrated in \textcolor{black}{Figure\,\ref{fig:accretion}}. We assume the velocity in the sky plane to be toward the galaxy disk \textcolor{black}{(i.e., radially inward)} and the same as our measured line-of-sight velocity. 
\begin{subequations}
\begin{equation}\label{eq:accretion}
\begin{split}
    {\dot{\rm  M}_{\rm CGM}}(r) = {\rm \int_A\int_{\Delta v_{emit}}} m_p n(r) \;\rm dv dA \\
    {\rm where}\; n(r) = \frac{{\rm N(HI)_{CGM}}(r)}{2r_s} 
\end{split}
\end{equation}
Here the radial profile of $\rm N(HI)_{CGM}$ comes from equation\,\ref{eq:density}. For the major axes pointings, we estimate the accretion rate also using the angular velocity, $\omega (r)$ in the disk plane:
\begin{equation}\label{eq:acc2}
\begin{split}
    \dot{\rm  M}_{\rm CGM}(r) =  {\rm M(HI)_{CGM}}(r)\omega(r) \\
    {\rm where}\; \omega(r) = {\rm \bar{v}_{CGM}}(r)/r 
\end{split}
\end{equation}
\end{subequations}

\begin{figure*}
    \centering
    \includegraphics[width=0.49\textwidth]{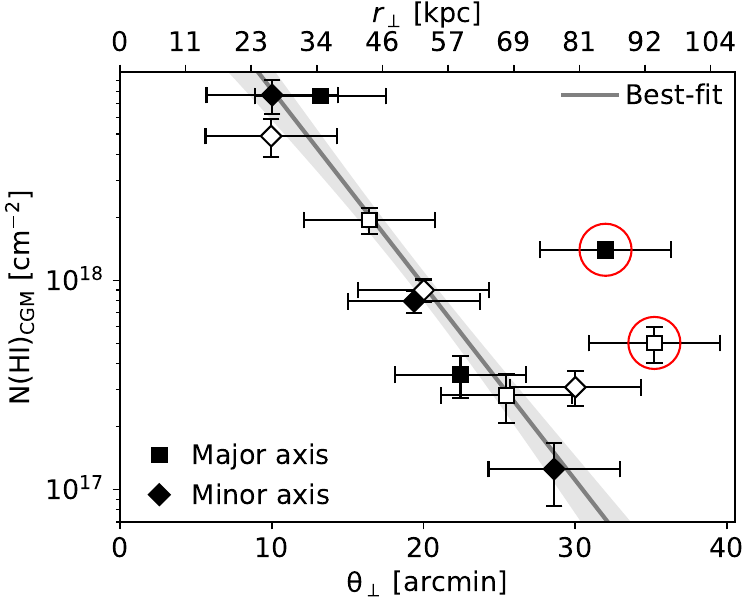}
    \includegraphics[width=0.49\textwidth]{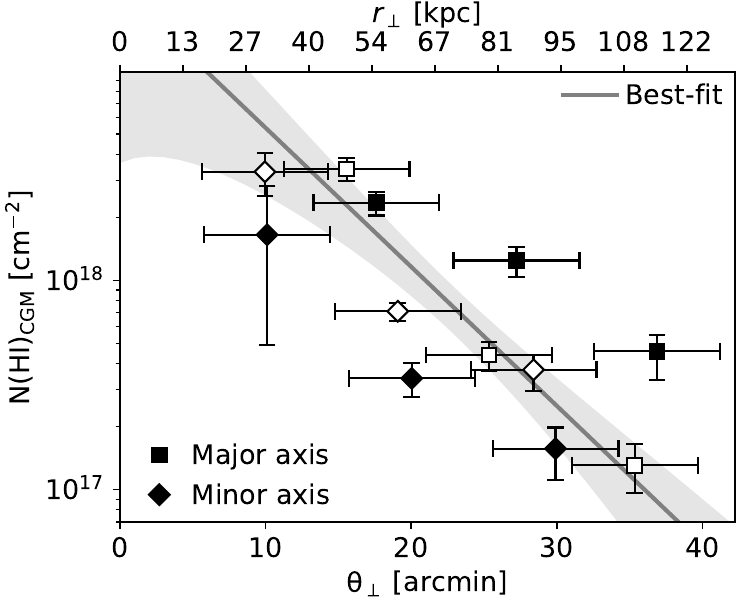}
    \caption{N(\hin) in the CGM of NGC\,891 (left) and NGC\,4565 (right), with best-fit models. The filled and unfilled symbols correspond to pointings at higher and lower declinations than the galaxy disk. The physical separation corresponding to the angular separation from the disks is noted in the top x-axis, adopting a distance of 9.2\,Mpc and 10.8\,Mpc for NGC\,891 and NGC\,4565 from us, respectively \citep{Heald2012}. In the left panel, the points marked within red circles have been excluded from the fit. NGC\,891 has a steeper profile than NGC\,4565, with more \hi closer to the disk of NGC\,891.} 
    \label{fig:CGM}
\end{figure*}

\section{Results and discussion}\label{sec:discuss}
Below we discuss our results on the density, mass, velocity, and accretion rate of the detected \hi in the CGM of NGC\,891 and NGC\,4565 and interpret them in the context of star-forming activities in these galaxies.

\subsection{Density profile}\label{sec:density}
We note the N(\hin)$\rm_{CGM}$ at each pointing in Table\,\ref{tab:results}. We show the best-fitted radial profile of N(\hin) in the CGM of NGC\,891 and NGC\,4565 in Figure\,\ref{fig:CGM}. 

In the CGM of NGC\,891 we have excluded two pointings, UP\,3J and DN\,3J, from the fit (highlighted with red circles in Figure\,\ref{fig:CGM}, left panel). These pointings are the farthest from the center of NGC\,891 along the major axes. Because the velocity of the \hi emission at UP\,3J pointing is significantly different from its adjacent pointings (see Figure\,\ref{fig:spectra891}), this emission might not be co-spatial with its adjacent pointings. Despite moving at similar velocities, the N(\hin) toward DN\,3J is larger than that toward DN\,2J, which is not suitable for an exponentially decreasing radial profile. The excess N(\hin) toward DN\,3J than DN\,2J might be a result of unresolved \hi cloud(s) embedded in diffuse \hin. Or, the radial profile of the diffuse \hi column density at this radius might become flatter than an exponential. Verifying this scenario would require further deep and high-spatial-resolution observation around DN\,2J--DN\,3J pointings. 

\begin{figure*}
    \centering
    \begin{subfigure}{0.39\textwidth}
        \includegraphics[width=\textwidth]{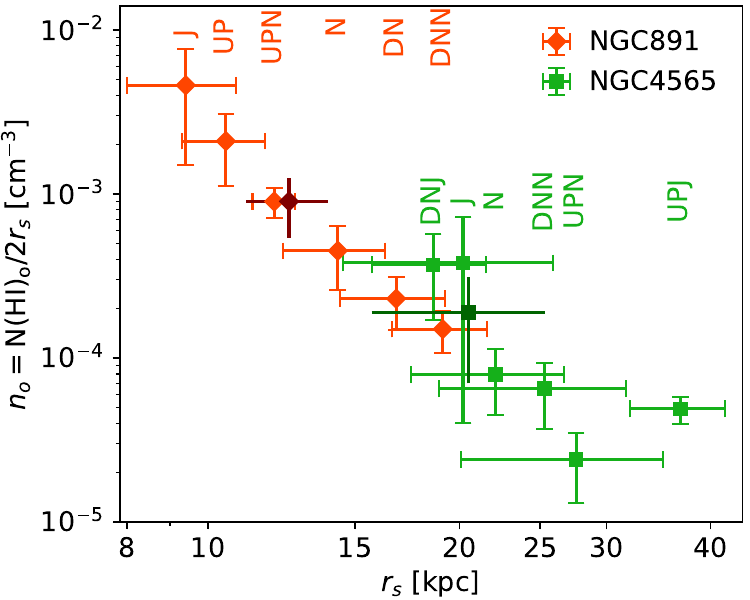}
        \caption{}
    \end{subfigure}
    \begin{subfigure}{0.6\textwidth}
        \includegraphics[width= \textwidth]{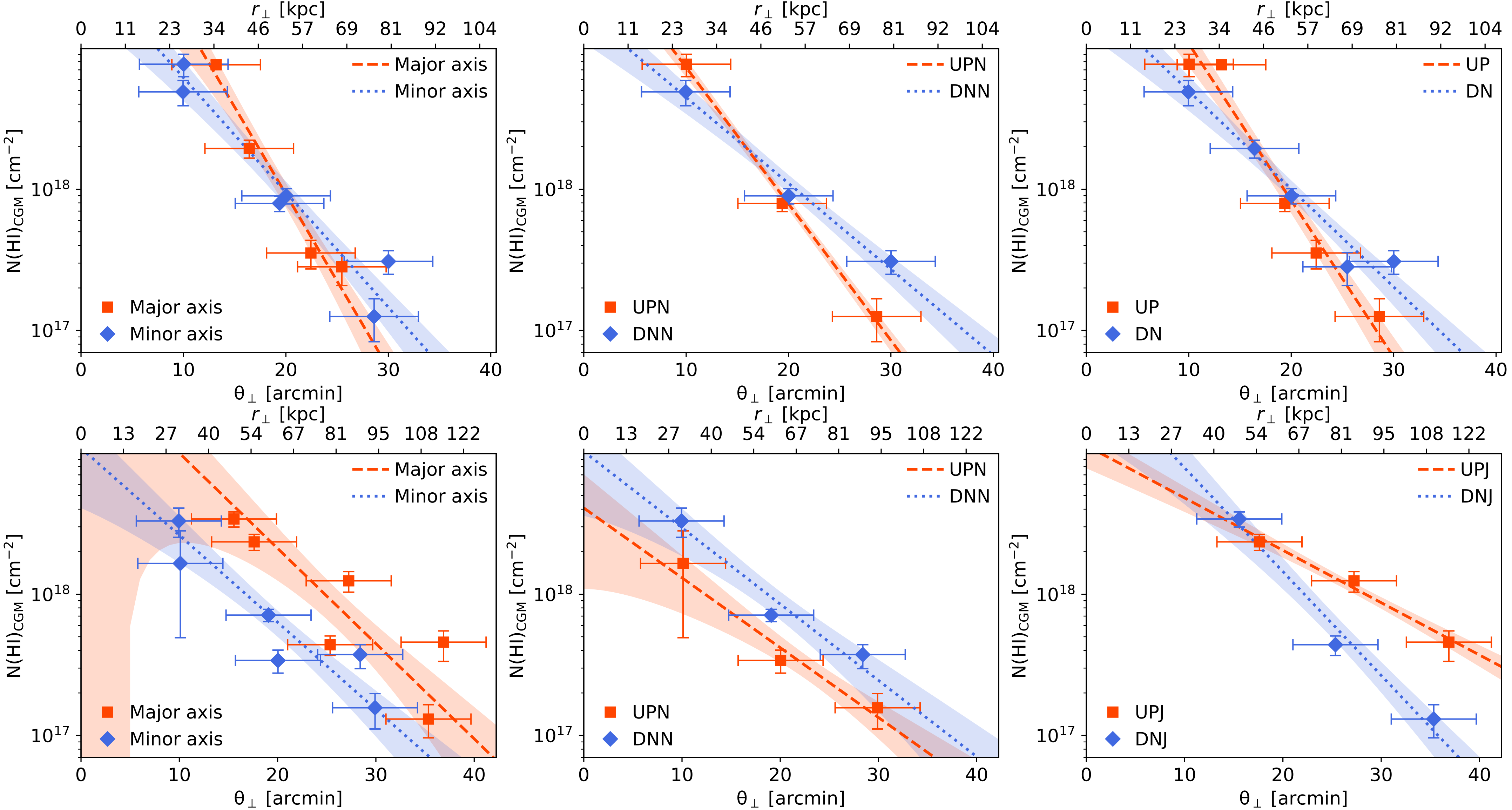}
    \caption{}
    \end{subfigure}
    \caption{\textbf{(a)} The \hi density at the center, $n_o =$N(\hin)$_o/2r_s$ vs the scale radius, $r_s$, for the models best-fitted to different subsets of the data, labeled accordingly. See Table\,\ref{tab:obs} for the terminology of the pointings. The markers with darker shades and uncapped error bars correspond to the global best fits (shown in Figure\,\ref{fig:CGM}). \textbf{(b)} N(\hin) in the CGM of NGC\,891 (top 3 panels) and NGC\,4565 (bottom 3 panels), with best-fit models for different subsets of the data, labeled in the legend of each panel.}
    \label{fig:fit}
\end{figure*}

Overall, NGC\,891 has a $1.6\pm 0.5\times$ steeper profile with $4.7_{-3.0}^{+13.3}\times$ higher central \hi density, $n\rm_o =$N(\hin)${\rm _o}/2r_s$, than NGC\,4565. It results in $1.6^{+9.1}_{-1.0}$ times more \hi in the extraplanar region (i.e., at and around UP/DN\,1 pointings) of NGC\,891 than NGC\,4565. This ratio is similar to the ratio of SFR of these galaxies. This supports the ``bathtub'' model of star formation where the main factor controlling the SFR of a galaxy is the available \hi in and around the galaxy disk \citep{Dekel2014}. Our result is qualitatively consistent with the finding of \citetalias{Das2020b} based on the four pointing (blue triangles in Figure\,\ref{fig:spectra891} and \ref{fig:spectra4565}) along the minor axes of these galaxies. 

In Figure\,\ref{fig:fit}, we show the best-fits (panel b) and the best-fit parameters (panel a) for the different subsets of our data: pointings only along the major axes (`J') or minor axes (`N'), only at higher declination (`UP') or lower declination (`DN') than the \hi disk, etc. In each galaxy, the scatter of $n\rm _o$ across $\approx$2 orders-of-magnitude and scatter of $r_s$ across a factor $\approx 3$ for these different subsets indicate that the circumgalactic \hi is not symmetric in two sides of the disk and along the azimuth. 

\subsubsection{NGC\,891}
\hi profiles are different along the major and minor axes, with the major axis having a $10.2_{-4.4}^{+1.8} \times$  denser and $1.5_{-0.4}^{+0.5} \times$ steeper \hi profile (Figure\,\ref{fig:fit}(b), top left). It could be due to the active accretion of \hi from the CGM to the disk along the major axis. 

The \hi profiles along the minor axis are different on two sides of the disk (Figure\,\ref{fig:fit}(b), top middle), with the higher declination side of the disk (`UPN') having a $6.0_{-2.3}^{+3.6}\times$ denser and $1.6\pm 0.3\times$ steeper \hi profile. There is already a galactic fountain along the minor axis \citep{Oosterloo2007,Fraternali2008}. Thus the \hi asymmetry in two sides of the disk along the minor axis might be due to a diffuse extended part of the fountain which is not mapped in shallow WSRT observations. 

We also fit the major and minor axes pointings at higher declination, i.e., UP\,J and UP\,N together, and compare it with lower declination pointings  (Figure\,\ref{fig:fit}(b), top right panel). It essentially splits the data into two diagonal sides of the disk, northwest and southeast, respectively. The \hi profile is $9.1_{-5.5}^{+11.7}\times$ denser and $1.6_{-0.4}^{+0.5}\times$ steeper at the higher declination (`UP') pointings. There is a \hi filament \citep{Oosterloo2007} toward UGC\,1807 in the northwest of NGC\,891 (green contours in the DSS image of Figure\,\ref{fig:spectra891}). Thus the \hi asymmetry in two diagonal sides of the disk might be the result of a diffuse extended arm of that filament which is not captured in shallow interferometric maps. 

\paragraph{What is the origin of \hi at UP3J?}
Given the shape and width of the \hi spectrum and the mass of the \hi emission at the UP\,3J pointing of NGC\,891 (top left panel of Figure\,\ref{fig:spectra891}), we cannot rule out the possibility of a dwarf galaxy. However, we did not find any known object within 10$'$ of this pointing at the similar line-of-sight velocity of the detected \hi emission in NED\footnote{\url{https://ned.ipac.caltech.edu/conesearch}}. Its velocity offset from the systemic velocity of NGC\,891 is similar to the high-velocity clouds found in the CGM of Milky\,Way \citep{Putman2012,Richter2017}. Whether or not the \hi emission detected at this pointing is a high velocity \hi cloud in the CGM of NGC\,891 and/or a yet undetected dwarf galaxy needs further investigation with deep interferometric and optical surveys, respectively.  

\subsubsection{NGC\,4565}
Along the major and minor axes, the \hi profiles have similar slopes (Figure\,\ref{fig:fit}(b), bottom left). The best-fit $n_o$ is higher by a factor of 4.8 along the major axis possibly due to active accretion, but it is consistent with $n_o$ along the minor axis due to the large errors in density estimate along the major axis (Figure\,\ref{fig:fit}(a)). 

The lower declination side of the disk along the major axis (`DNJ') has a $7.6_{-4.6}^{+6.8}\times$ denser and $2.0_{-0.5}^{+0.7} \times$ steeper \hi profile than the higher declination side (Figure\,\ref{fig:fit}(b), bottom right). The asymmetry might have caused/be caused by the strong warp around the disk of NGC\,4565 observed in 21-cm \citep{Zschaechner2012}.
Also, NGC\,4565 hosts $\sim$5-6 dwarf galaxies \citep{Carlsten2020} along the higher declination arm (`UPJ'). The flatness in the \hi profile on this side might be due to ram-pressure/tidally stripped \hi from these dwarf satellite galaxies. 

Along the minor axis, the \hi profiles have similar slopes on both sides of the disk, but $n_o$ is $2.7_{-1.6}^{+4.4}\times$ larger on the lower declination side of the disk (`DNN'), resulting in more \hi mass in the CGM (Figure\,\ref{fig:fit}(b), bottom middle). This enhancement could be due to tidal interactions between the \hi disks of NGC\,4565 and its companion NGC\,4562 to the southwest (contours in the DSS image of Figure\,\ref{fig:spectra4565}).

Thus, the detected circumgalactic \hi can be primarily explained by a diffuse extended \hin, with some \hi clouds embedded in it, and the interaction of the target galaxies with their companions.  

\subsection{Mass}\label{sec:mass}
\begin{figure*}
    \centering
    \includegraphics[width=0.49\textwidth]{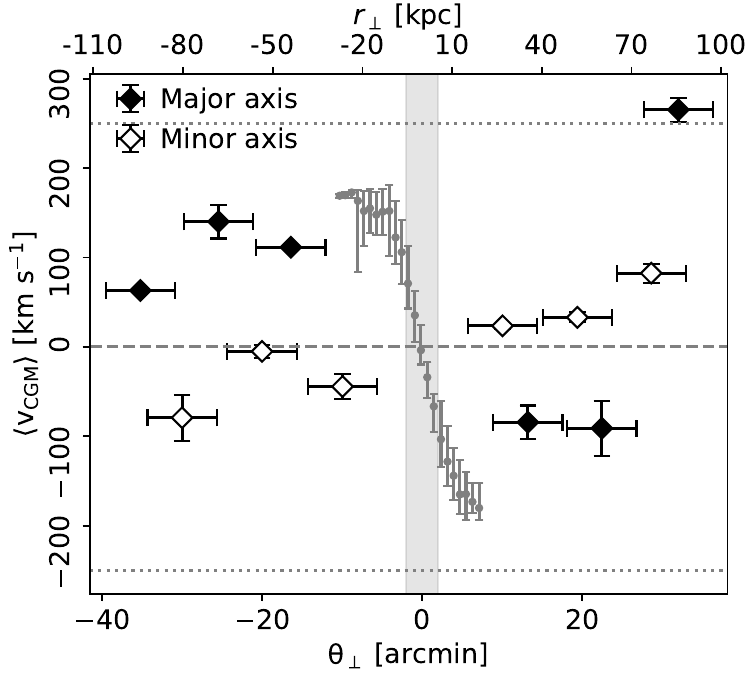}
    \includegraphics[width=0.49\textwidth]{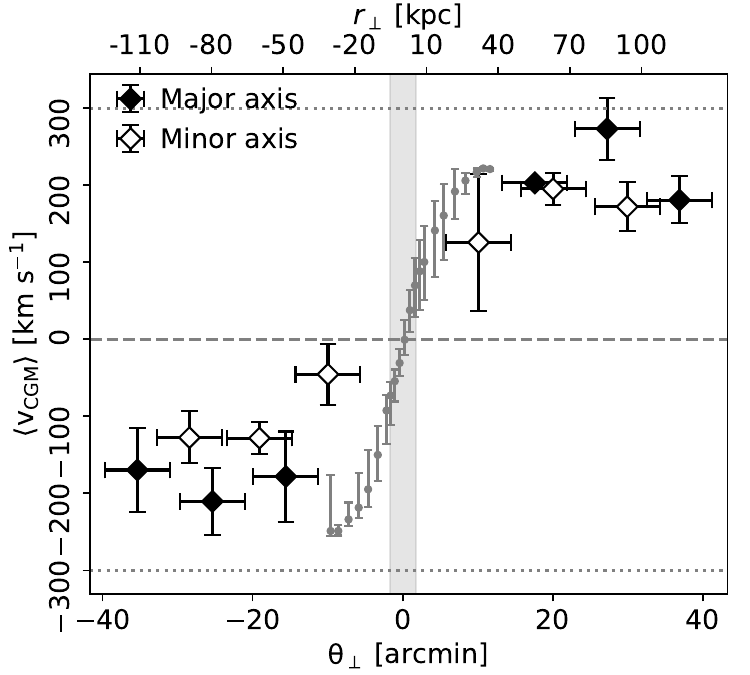}
    \caption{The line-of-sight mean velocity of \hi in the CGM of NGC\,891 (left) and NGC\,4565 (right) in the rest frame of the galaxy. The horizontal dotted lines denote the maximum rotational velocity of the \hi disk measured in the GBT data. The gray circles show the rotation curve of the disk along the major axis, measured in the WSRT data. The spatial extent of the disk along the minor axis is shown with the gray vertical patch. The horizontal dashed line is drawn at 0 km\,s$^{-1}$ to guide the eye.}
\label{fig:v}
\end{figure*}
We note the M(\hin)$\rm_{CGM}$ at each pointing within the beam in Table\,\ref{tab:results}. For NGC\,891 (excluding the UP3J/DN3J pointings), M(\hin)$\rm_{CGM}$ along and around (within $\pm\pi/4$) the major and minor axes are $3.2 \pm 1.7 \times 10^8$\msun~and $1.4 \pm 0.2 \times 10^8$\msun, respectively, resulting in total mass of $4.6 \pm 1.7 \times 10^8$\msun. For NGC\,4565, M(\hin)$\rm_{CGM}$ along and around the major and minor axes are $2.9 \pm 0.6 \times 10^8$\msun~and $1.0 \pm 0.2 \times 10^8$\msun, respectively, resulting in the total mass of $3.9 \pm 0.6 \times 10^8$\msun. 

Thus, both galaxies have more mass along the major axes than the minor axes, indicating active accretion along the major axes. The masses quoted in \citetalias{Das2020b} based on minor axes pointings and assuming azimuthal symmetry were underestimated. 

M(\hin)$\rm_{CGM}$ is $11.4\pm 4.2$\% and $5.4\pm0.8$\% of the M(\hin)$\rm_{disk}$ of NGC\,891 and NGC\,4565, respectively. Based on the GBT and WSRT measurements within 50\,kpc of NGC\,891 and NGC\,4565, \citetalias{Pingel2018} estimated $f_{19}$, the fraction of \hi mass below N(\hin) = 10$^{19}$ cm$^{-2}$ to be 0.4--0.6\% and 0.7--0.9\%, respectively. Our measurement is larger than the estimation of \citetalias{Pingel2018} by an order of magnitude. It suggests the presence of a large amount of diffuse, extended \hi with N(\hin) below $10^{19}$ cm$^{-2}$, extending beyond 50\,kpc. This emission might have been missed due to the lower sensitivity of the GBT observations presented in \citetalias{Pingel2018}. Also, $f_{19}$ in \citetalias{Pingel2018} was based on azimuthal averages in circular annuli around the galaxies, whereas we focus on pointings along principal axes. Therefore, most of the circumgalactic \hi could be concentrated along principal axes of the galaxies, leading to a higher estimate in our case.

\begin{figure}
    \centering
    \begin{subfigure}{0.235\textwidth}
    \includegraphics[width=\textwidth]{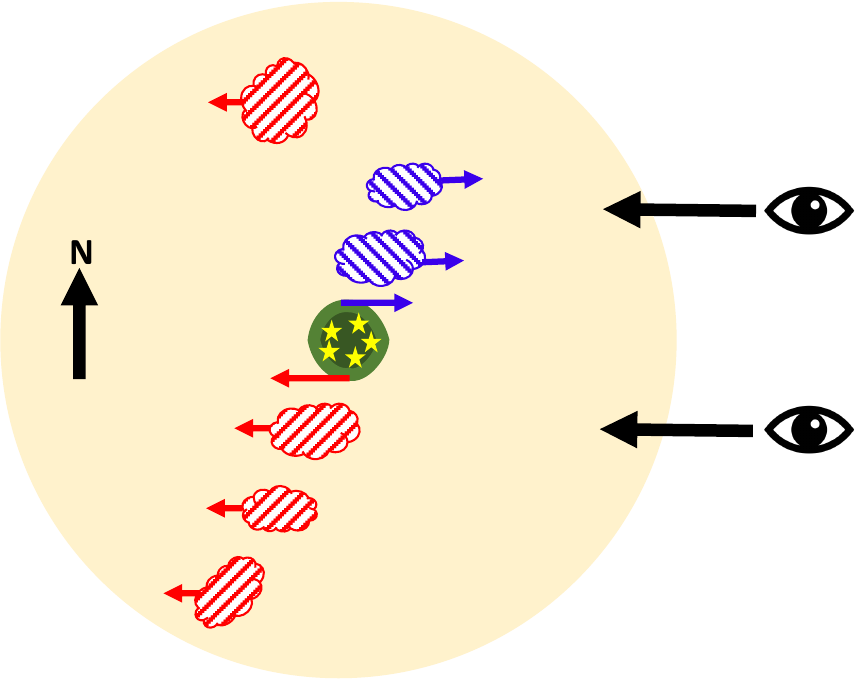}
    \caption{}
    \end{subfigure}
    \begin{subfigure}{0.235\textwidth}
    \includegraphics[width=\textwidth]{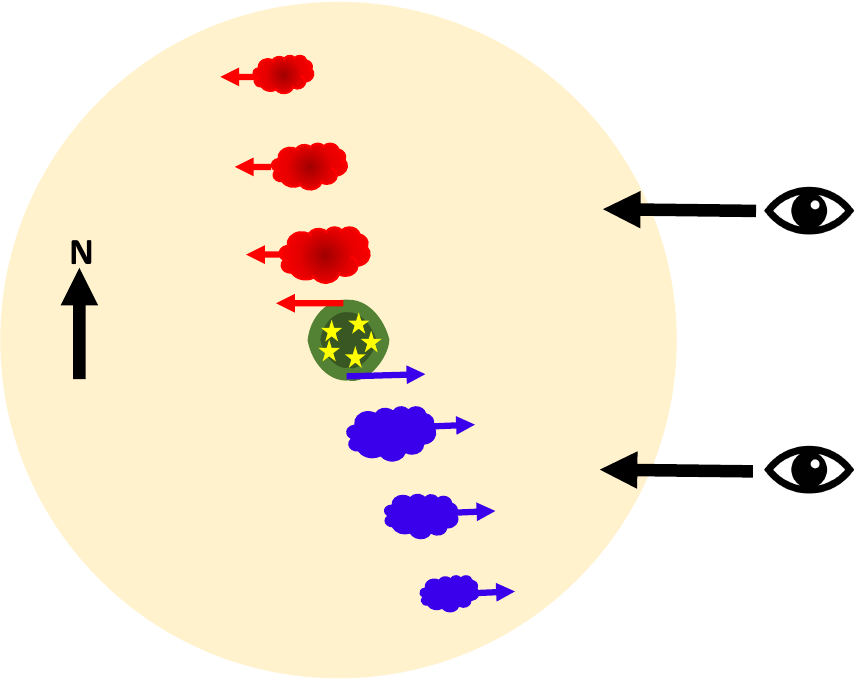}
    \caption{}
    \end{subfigure}
    \begin{subfigure}{0.235\textwidth}
    \includegraphics[width=\textwidth]{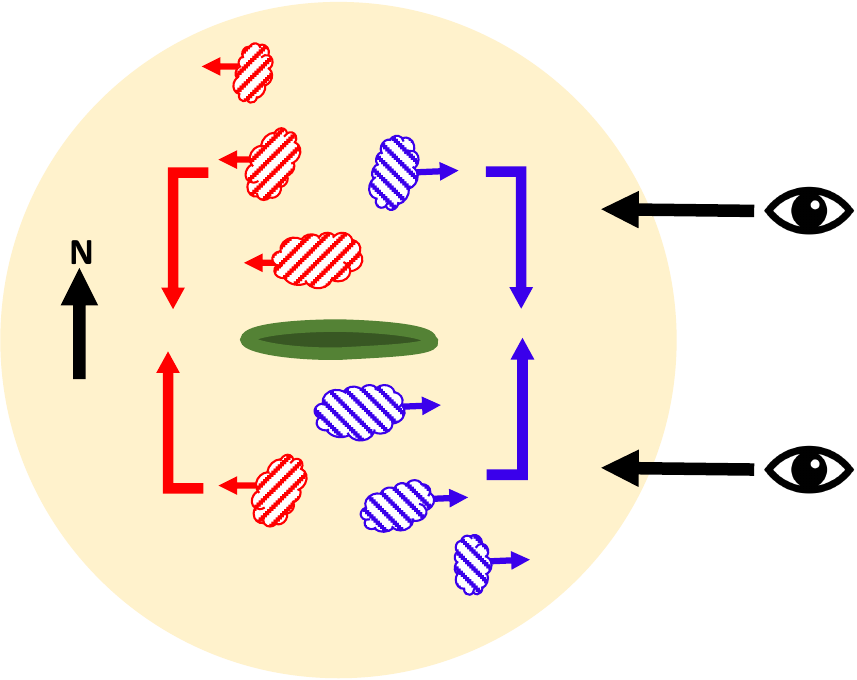}
    \caption{}
    \end{subfigure}
    \begin{subfigure}{0.235\textwidth}
    \includegraphics[width=\textwidth]{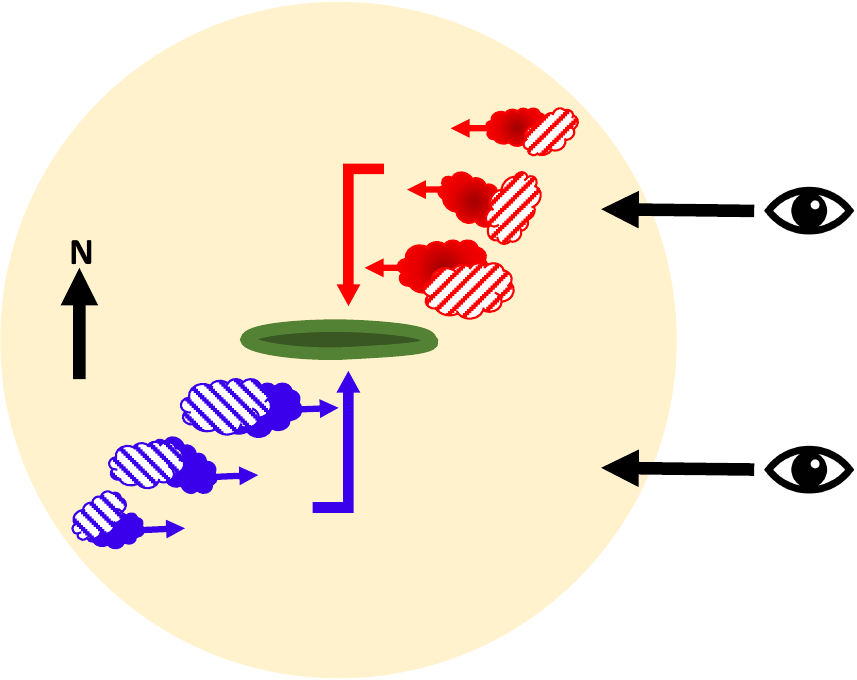}
    \caption{}
    \end{subfigure}
    \caption{The location of the circumgalactic \hi along the line-of-sight that we assume while calculating the accretion rate. Blue-shifted and red-shifted emissions are shown in blue and red. \hi in the CGM of NGC\,891 and NGC\,4565 are shown with hatched and filled clouds. The disk, the halo, and the distance of clouds from the disk are not scaled, the size of clouds is not to scale with column densities, neither do we claim the \hi to be composed of only clouds and not extended structures; it is only for illustration. The pointings along the major axes are shown in panels (a) and (b). The bottom two panels show the pointings along the minor axes with possible scenarios of a fountain (c) or inflow (d).}
    \label{fig:accretion}
\end{figure}
The depletion timescale, $\rm \tau_{dep} =$ (M(\hin)$\rm_{disk}$ + M(\hin)$\rm_{CGM})/SFR$ is $2.1\pm0.8$\,Gyr for NGC\,891 and $11.4\pm 1.8$\,Gyr for NGC\,4565. By combining this information with the amount of \hi in the extraplanar region estimated in the previous subsection, we infer the following. NGC\,891 might have accreted most of its star-forming fuel by now and will soon run out of fuel unless there is a huge \hi reservoir in the regions not probed in our observations -- regions away from the principal axes, and regions at the larger galactocentric radius. NGC\,4565, on the other hand, is forming stars almost quiescently; it has an ample supply of \hi to continue forming stars at this rate. 

\subsection{Velocity}\label{sec:velocity}
We show the line-of-sight mean velocity of the circumgalactic \hi in Figure\,\ref{fig:v} and quote them in Table\,\ref{tab:results}. Except for the UP\,3J pointing of NGC\,891, the velocities at all other pointings of both galaxies display interesting trends that we discuss below. 

\subsubsection{Major axis}
In both galaxies, the \hi velocity in the CGM along the major axes is aligned with the disk rotation (illustrated in panels (a) and (b) of Figure\,\ref{fig:accretion}). Such co-rotation has been observed previously in a few \mgii absorbers at $z\approx 0.2$ \citep{Ho2017,Martin2019} and at $z\approx 1$ \citep{Zabl2019} and in one Ly$\alpha$ absorber at $z\approx 0.6057$ \citep{Weng2023}, but not in \hi emission at $z\approx 0$. It could be the neutral CGM that is accreting onto the disk keeping the angular momentum conserved. 

Otherwise, the \hi disk might not truncate sharply and could be larger than what is found in shallow ($\approx 10^{19}\;\rm cm^{-2}$) interferometric observations (the green contours on DSS images in Figure\,\ref{fig:spectra891} and \ref{fig:spectra4565}). In that case, our detected \hi would be the outer part of an extended \hi disk. The column density of this co-rotating \hi is consistent with the predicted ``proto-disk'' extended out to $\approx$80\,kpc, modulo the degeneracy between the volume-filling factor, covering fraction and intensity of the cosmic ionizing background \citep{Bland-Hawthorn2017}.

Distinguishing between the scenarios of a co-rotating neutral CGM or an extended \hi disk is beyond the scope of this paper. It would require deep co-spatial observations of H$\alpha$ emission and \hi emission at high angular resolution in the future. 

\subsubsection{Minor axis}
The velocities of circumgalactic \hi along the minor axes show a similar gradient in both galaxies (Figure\,\ref{fig:v}), although the velocities are smaller in NGC\,891 than NGC\,4565. 

Here the velocity profile is complicated by the possibility of inflow, fountain, or outflow. The interpretation of a pure inflow or a pure outflow is degenerate to the location of the \hi emitter in the CGM along our line-of-sight. In an inflow, the redshifted/blueshifted emission is on the nearer/farther side of the disk than the observer (panel (d) of Figure\,\ref{fig:accretion}). The fountain, in principle, has both outflowing and inflowing components. If the fountain is not perfectly aligned along the minor axis, the mean line-of-sight velocity would depend on whether it is titled toward/away from the observer. If the redshifted/blueshifted emission is on the farther/nearer side of the disk, it would be an outflow or fountain if its velocity in the sky plane is away from or toward the disk, respectively. 

In NGC\,891, the presence of a galactic fountain is already known from the increasing lag in rotation with height \citep[also discussed in \S\ref{sec:density}]{Oosterloo2007,Fraternali2008}. Our observed velocity gradient could be due to a cosmic inflow and/or a diffuse extended part of the fountain that is tilted away from us. In the latter scenario, the smaller velocities compared to the maximum rotational velocity of the disk could be due to \hi emitters moving at opposite velocities in the fountain suppressing the mean velocity (panel (c) of Figure\,\ref{fig:accretion}). 

NGC\,4565 hosts a low-luminosity type\,2 AGN \citep{Veroncetty2000}, which could be responsible for a nuclear outflow. Also, based on 144\,MHz observations tracing the non-thermal emission from cosmic ray electrons in the extraplanar region of NGC\,4565, \cite{Heesen2019} argue that NGC\,4565 could be in transition from a weak outflow-dominated phase to an inflow-dominated phase in the past 40\,Myr. Thus we cannot conclusively comment on the origin of the \hi emitters along the minor axis of NGC\,4565 without constraining their velocity in the sky plane. While inflows and outflows should have distinct metallicity, measuring it might not be useful in reality as the CGM is predicted to be inhomogeneously mixed in simulations and there is no observational evidence for a correlation between absorber metallicity and azimuthal angle \citep{Martin2019,Weng2023}.   

\subsection{Accretion rate}\label{sec:accretion}
In NGC\,891, the total accretion rate, $\dot{\rm  M}_{\rm CGM}$, \textcolor{black}{at the position of all observed pointings integrated within the GBT beam} is $0.30\pm 0.03 \rm \,M_\odot yr^{-1}$. For all pointings except UP3J and DN3J (as they were excluded from the fitting of N(\hin)$\rm_{CGM}$ in \S\ref{sec:density}), we extend \textcolor{black}{the accretion rate calculation} to each quadrant in the CGM (i.e., $\pm\pi/4$ region around the major and minor axes) assuming azimuthal symmetry, and obtain $\dot{\rm  M}_{\rm CGM}$ of $0.73\pm 0.08 \rm \,M_\odot yr^{-1}$. Adding it to the accretion rate measured within the GBT beam of the excluded pointings, the total $\dot{\rm  M}_{\rm CGM}$ is $0.80\pm 0.08 \rm \,M_\odot yr^{-1}$. $\dot{\rm  M}_{\rm CGM}$ along the major and minor axes are $0.65\pm 0.08 \rm \,M_\odot yr^{-1}$ and $0.15\pm 0.01 \rm \,M_\odot yr^{-1}$. 

In NGC\,4565, the total $\dot{\rm  M}_{\rm CGM}$ \textcolor{black}{at the position of all observed pointings integrated within the GBT beam} is $0.20\pm 0.02 \rm \,M_\odot yr^{-1}$. Extending \textcolor{black}{the accretion rate calculation} to each quadrant in the halo, we obtain a total $\dot{\rm  M}_{\rm CGM}$ of $0.83\pm 0.08 \rm \,M_\odot yr^{-1}$. $\dot{\rm  M}_{\rm CGM}$ along the major and minor axes are  $0.66\pm 0.07 \rm \,M_\odot yr^{-1}$ and $0.17\pm 0.03 \rm\; M_\odot yr^{-1}$.

Using the second definition of accretion rate (equation\,\ref{eq:acc2}), $\dot{\rm  M}_{\rm CGM}$ along the major axes of NGC\,891 and NGC\,4565 are $0.29\pm 0.04 \rm \,M_\odot yr^{-1}$ and $0.57\pm 0.06 \rm \,M_\odot yr^{-1}$, respectively. 

Thus, the total accretion rate in both galaxies is similar, and it is larger along the major axes (irrespective of the definition of accretion rate used) than the minor axes, indicating active accretion onto the disk. The specific star formation rate, sSFR = SFR/M$_\star$, and surface density of SFR, $\rm \Sigma_{SFR}$ are an order-of-magnitude larger in NGC\,891 than NGC\,4565 (see Table\,\ref{tab:results}). Despite this difference, the similarity in \hi accretion rate suggests that the intermediate stages between the accumulation of \hi and star-formation, e.g., the \hi--$\rm H_2$ conversion, are inefficient in NGC\,4565 compared to NGC\,891.

\subsubsection{Tracing accretion: found or missing?}
In NGC\,4565, $\dot{\rm  M}\rm _{CGM}/SFR$ is $1.24\pm0.12$. Because we cannot distinguish between an outflow or an inflow along the minor axis of NGC\,4565, excluding the estimates along the minor axis, we obtain $\dot{\rm  M}\rm _{CGM}/SFR$ of $0.98\pm0.10$ (or $0.85\pm0.09$ using the second definition of accretion rate in equation\,\ref{eq:acc2}). This is arguably the first galaxy where the accretion rate is consistent with the SFR. It implies that the CGM of NGC\,4565 is continuously supplying \hi to the disk for a sustainable star formation. 
Excluding NGC\,4562 and IC\,3571 that are already detected in WSRT maps, NGC\,4565 hosts 19 dwarf galaxies down to $M_g \sim -10$ within 150\,kpc \citep{Carlsten2020}, with \hi masses of each being $< 10^{6.88}$\msun \citep{Karunakaran2022}. Determining if the accreting \hi is dominated by the cosmic inflow or the ram-pressure/tidally stripped \hi from these satellite galaxies is beyond the scope of this paper. If \hi along the minor axis pointings trace an outflow, the mass-loading factor, $\dot{\rm M}/\rm SFR$, is $0.25\pm0.04$. 

In NGC\,891, $\dot{\rm  M}\rm _{CGM}/SFR$ is $0.36\pm0.04$ (or $0.13\pm0.02$ using the second definition of accretion rate in equation\,\ref{eq:acc2}). Thus, $\approx$60--90\% of the required accreting material is still missing. Below we discuss some of the possibilities of where the missing accretion could be:
\begin{enumerate}[wide =0 pt,itemsep=-0.1em]
\item \textbf{Beam smearing:} 
If the length scale of the \hi in the sky plane is smaller than the GBT beam,
its intensity could be underestimated due to the GBT beam smearing that dilutes the intensity of any probed structure. The smaller the size of the \hi emitters, the higher the dilution.  
The WSRT can capture up to the length scale of $20'$ which is larger than the GBT beam. Thus the beam smearing is not expected to affect the excess \hi detected by the GBT than the WSRT.  
However, our GBT observation is more than 2 orders-of-magnitude deeper than the WSRT maps. Thus, some of the excess \hi detected by the GBT could be smaller than the GBT beam whose column density is below the sensitivity of the WSRT maps. While we can rule it out for pointings close to the disk (Appendix\,\ref{sec:intermediate_convol}), it remains plausible for the far-away pointings. 
\item \textbf{Opacity:} If the length scale of the \hi along the line of sight is smaller than that assumed here, the number density and hence the opacity of the measured \hi would be higher. Thus, for a given intensity, the column density would increase after correcting for the self-shielding. This is more likely to be the case for the major axis pointings if the corotating \hi we find in \S\ref{sec:velocity} is genuinely the outermost disk instead of the CGM. 
\item \textbf{Sky-plane motion:} We have calculated the accretion rate based on the line-of-sight motion. If the circumgalactic \hi is \textcolor{black}{spiraling in with the velocity in the sky plane being larger than the line-of-sight velocity}, the accretion rate could be larger.  
\item \textbf{Ionization:} Given the mass of NGC\,891, it is likely a ``hot mode'' accretion where the \hi forms out of the hot CGM. Near the galaxy disk, \hi is pressure confined due to the larger density and temperature of the hot CGM. At the interface of the hot ionized CGM and the neutral CGM, partially ionized warm/cool CGM can form because of thermal conduction and enhanced cooling rate due to condensation \citep{Li2020}. If our detected \hi is partially ionized, the mass of \hi containing gas would be higher than the current estimate. The hydrogen is rapidly ionized at $\approx 10^4$\,K, thus a 60--90\% ionization is not impossible.
\item \textbf{Off-axis:} If most of the accretion is happening away from the principal axes, the accretion rate might have been underestimated. 
\item \textbf{Larger distance:} There might be more \hin, diffuse and/or compact, at larger impact parameters. The \hi is less likely to be destructed at large galactocentric distances due to the lower density of the surrounding hot CGM \citep{Li2020}. Whether or not it immediately contributes to the SFR would depend on its density and velocity. 
\end{enumerate}

Testing these possibilities would require highly sensitive mapping of the whole CGM out to the virial radii of these galaxies using single-dish and interferometers in the future. 
\subsubsection{In-situ fueling}
Most of the discussion has assumed that the circumgalactic \hi contributes to the star formation in the disk. However, if there is in-situ star formation in the CGM of these galaxies, the circumgalactic \hi might be locally supplying fuel to those star formation hubs. 

NGC\,891 has abundant stellar structures in its CGM, e.g., a thick star-forming cocoon extended out to 40\,kpc/15\,kpc along the major/minor axis, a giant $\sim$90\,kpc stellar stream, a few arching stellar streams up to $\sim$50\,kpc looping around the stellar disk and a stellar halo extended out to $\sim$50\,kpc \citep{Mouhcine2010,Monachesi2016}. NGC\,4565 has a $\sim$60\,kpc stellar halo, but it is more massive, flatter, and is a larger fraction of the stellar disk mass in NGC\,891 than in NGC\,4565 \citep{Harmsen2017}. The stellar metallicity, [Fe/H], steeply declines radially in the stellar halo, implying a comparatively young stellar population in the CGM \citep{Monachesi2016}. NGC\,4565, showing no evidence of stellar streams/cocoons, has its stellar disk sharply truncated at 3\,kpc/26\,kpc along minor/major axis \citep{Martinez2019}. The inner CGM of NGC\,891 has an extended dust component likely embedded in an atomic gas \citep{Howk1997,Bocchio2016,Yoon2021} while NGC\,4565 does not have any \citep{Howk1999}. 
Also, NGC\,891 has only 7 (3 confirmed) dwarf satellite galaxies within 200\,kpc \citep{Carlsten2022}, while NGC\,4565 has 21 satellite galaxies within 150\,kpc \citep{Carlsten2020}. Whether some of the stellar structures in the CGM of NGC\,891 are disrupted satellites, and whether NGC\,891 might have accumulated \hi through minor mergers with its satellites in the past are not currently constrained. 

Thus, the \hi in the CGM of NGC\,891 and NGC\,4565 might have different origins. The circumgalactic \hi might be locally fueling star formation in the CGM of NGC\,891. In that scenario, the line-of-sight velocities of \hi would follow that of the stellar structures in the CGM. NGC\,4565, on the other hand, might be accreting through tidally/ram-pressure stripped \hi from its satellites. This might explain the different velocities of circumgalactic \hi in these galaxies.   

\section{Conclusions}\label{sec:conclude}
In this paper, we have extended the study of \cite{Das2020b} and presented the GBT observations of NGC\,891 and NGC\,4565 in 21-cm emission along their major axes out to $\sim 100-120$\,kpc in the CGM, and also revisited the observations in \citetalias{Das2020b}. Instead of the traditional mapping which is more expensive due to the required exposure time, we stare at a few selected (24) pointings for $\approx 3-4$\,hours and achieve an unprecedented $5\sigma$ sensitivity of $6.1\times 10^{16}$ cm$^{-2}$ for 20\,km\,s$^{-1}$ velocity width at 5\,km\,s$^{-1}$ velocity resolution. 

We substantially revised our GBT data reduction routine and the baseline subtraction method. We follow the methods outlined in \citetalias{Pingel2018,Das2020b} to compare our GBT observation and interferometric WSRT maps from the HALOGAS survey. We include the systematic uncertainty to account for the azimuthal asymmetry in the GBT beam and the variation in position angle throughout our GBT observation. We also rigorously test the angular size of the detected diffuse \hi by redefining noise and masking through intermediate convolution of the WSRT data cubes. Below we list our scientific results:
\begin{enumerate}[wide=0 pt,itemsep=-0.1em]
    \item We detect \hi emission at all GBT pointings at $>4.3\sigma$ significance including systematic uncertainties, with 3 pointings below $5\sigma$.
    \item More than $31-43$\% and $64-73\%$ of the \hi emission detected by the GBT at small and large impact parameters cannot be explained by the WSRT maps. It implies the presence of a diffuse neutral CGM. The shape, velocity width, and column density of the diffuse \hi are quite diverse, indicating a complex morphology and kinematics.
    \item The CGM of NGC\,891 has a steeper profile than NGC\,4565, with the ratio of \hi mass in the extraplanar region being $2^{+9}_{-1}$. It is consistent with their SFR ratio, supporting the \textcolor{black}{``bathtub''} model of star formation.  
    \item The N(\hin) profiles have different normalizations and scale radii for different subsets of the data, indicating a lack of azimuthal symmetry and axial symmetry around the \hi disk. These asymmetries can be attributed to the diffuse extended arm of the filament and fountain of NGC\,891, ram-pressure stripped \hi from the satellite galaxies of NGC\,4565, and tidal interaction with the companion galaxies of NGC\,891 and NGC\,4565.
    \item The mass ratio of \hi in the CGM and the disk is $0.11\pm 0.04$ and $0.05\pm0.01$ for NGC\,891 and NGC\,4565, an order of magnitude larger than previous estimates based on shallow ($\approx 10^{18}$ cm$^{-2}$) GBT mapping within 50\,kpc of these galaxies.  
    \item The velocity of the diffuse \hi along the major axes pointings are aligned with the disk rotation in both galaxies, suggesting a corotating CGM or the outer part of an extended \hi disk. Along the minor axes pointings, the velocities resemble the signatures of an inflow and/or a galactic fountain in NGC\,891. In NGC\,4565 we cannot distinguish between the scenarios of an inflow or an outflow.
    \item The depletion timescale and accretion rate in NGC\,4565 are $>9.6$\,Gyr and $>0.51$\,M$_\odot$\,yr$^{-1}$, large enough to sustain its star formation. NGC\,891, with $60-90\%$ of the required accreting material still missing, would run out of its raw fuel by $2.1\pm 0.8$\,Gyr. 
\end{enumerate}
The Parkes-IMAGINE (Imaging Galaxies Intergalactic and Nearby Environments) survey has mapped the entire CGM of 28 nearby galaxies in \hin, reaching sensitivities comparable to our GBT observation \citep{Sardone2020}. In the future, coordinated and deeper interferometric surveys and single-dish observations outfitted with phased array feeds or FLAG \citep[The Focal L-band Array for the GBT;][]{Pingel2021} architecture would substantially improve our understanding of the large-scale \hi accretion in the CGM, the \textit{true} radial extent of \hi disk, and the relation between \hi accretion and star-formation in the disk. Measurements of low ionized metal absorbers would constrain the gas-phase metallicity and the ionization fraction of the \hi emitters. Along with kinematic information, it would help disentangle the origin of circumgalactic \hi unambiguously and better understand the role of \hi in galaxy evolution. 

\section*{acknowledgments}
\textcolor{black}{We thank the anonymous referee for the constructive comments and suggestions.}
The Green Bank Observatory is a facility of the National Science Foundation operated under a cooperative agreement by Associated Universities, Inc. This research made use of data from Westerbork Synthesis Radio Telescope (WSRT) HALOGAS-DR1. The WSRT is operated by ASTRON (Netherlands Institute for Radio Astronomy) with support from the Netherlands Foundation for Scientific Research NWO. S.D. acknowledges support from the Presidential Graduate Fellowship from the Ohio State University, and KIPAC Fellowship from KIPAC, Stanford University. DJP acknowledges support from the South African Research Chairs Initiative of the Department of Science and Technology and the National Research Foundation. This research has made use of NASA's Astrophysics Data System Bibliographic Services.

\section*{Data Availability}
The raw GBT data presented here is publicly accessible in the NRAO archive\footnote{\url{https://data.nrao.edu/portal/}}. The WSRT data cubes are in the HALOGAS\footnote{\url{https://www.astron.nl/halogas/data.php}} website. The reduced and analyzed data are available from the corresponding author upon reasonable request. 

\bibliographystyle{mnras}


\appendix
\counterwithin{figure}{section}
\section{Baseline subtraction}\label{sec:baseline}
\begin{figure*}
    \centering
    \includegraphics[width=0.75\textwidth]{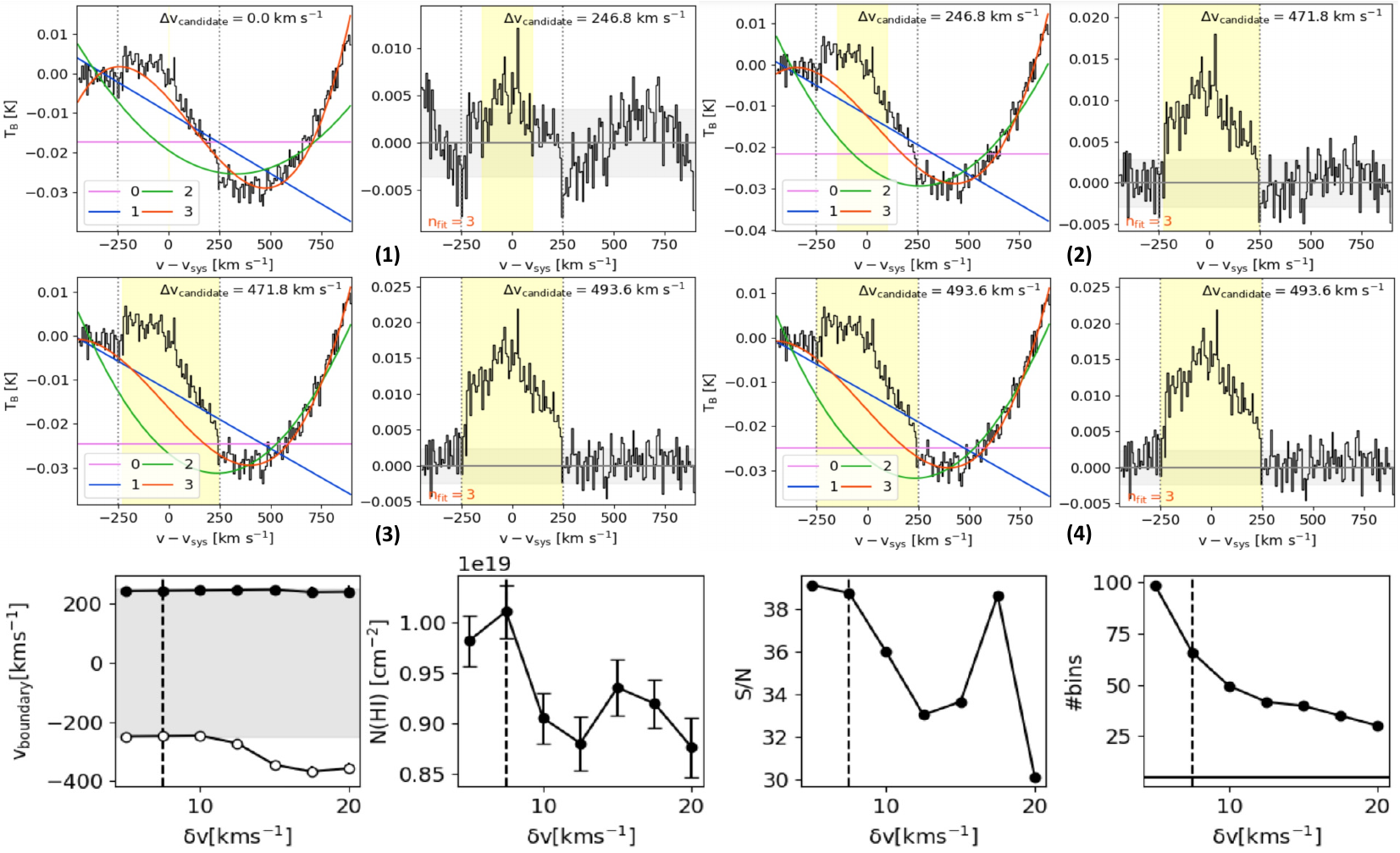}
    \includegraphics[width=0.75\textwidth]{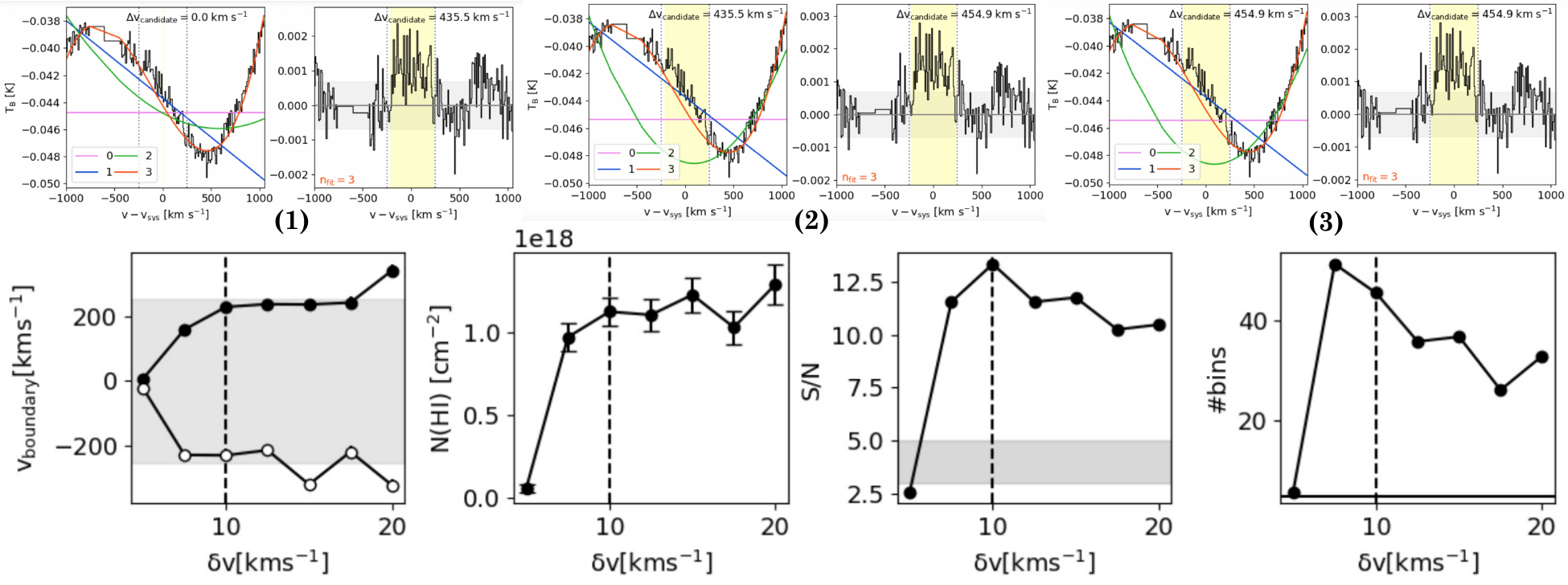}
    \includegraphics[width=0.75\textwidth]{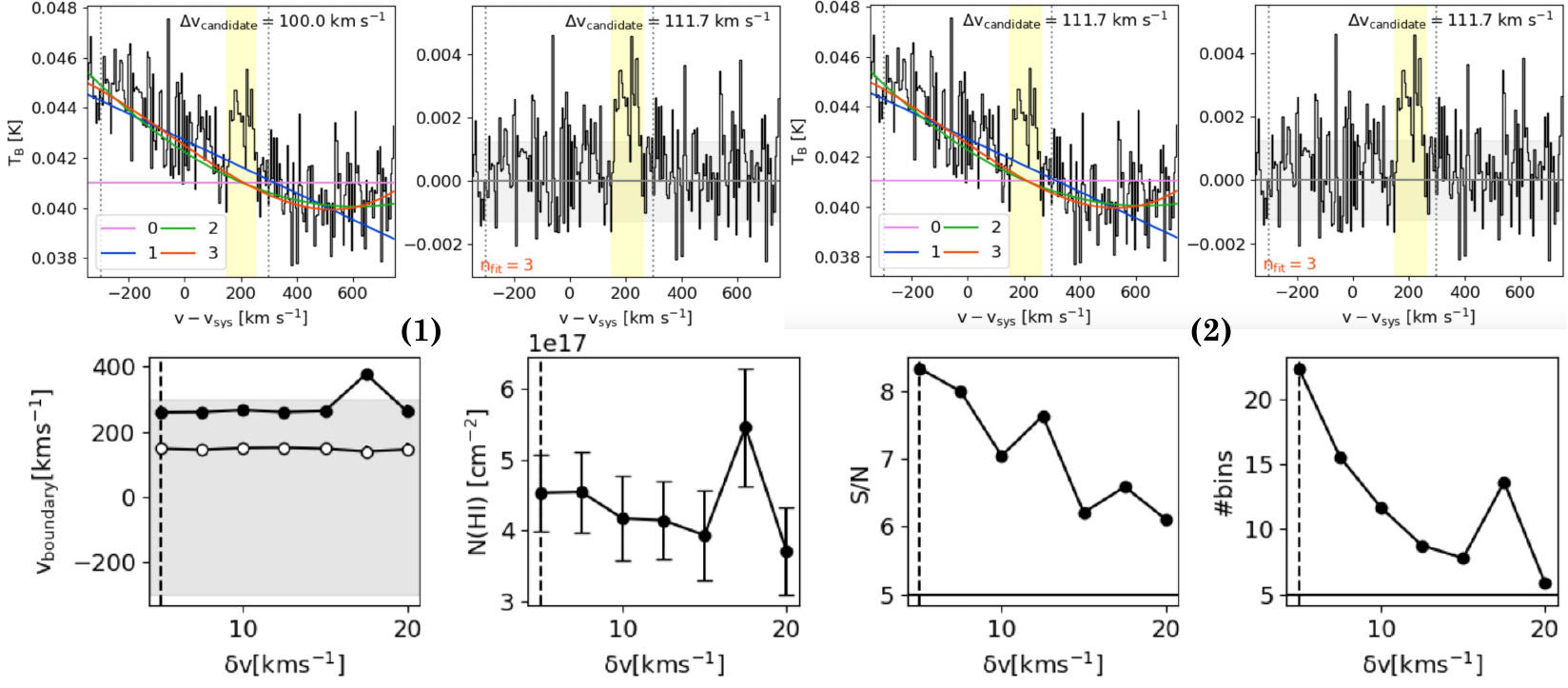}
    \caption{Illustration of the baseline fitting and subtraction routine. Top 3 rows: `DN1N' pointing of NGC\,891 for $\rm \delta v = 7.5\;km\;s^{-1}$, 4th and 5th rows: `DN2N' pointing of NGC\,891 for $\rm \delta v = 10\;km\;s^{-1}$, bottom 2 rows: `UP2N' pointing of NGC\,4565 for $\rm \delta v = 5\;km\;s^{-1}$. \textbf{Top 2 rows, 4th and 6th row:} Each iteration, noted in parentheses, consists of two panels. The left panels show the original spectrum with the candidate emission region indicated and the best-fit baselines of different polynomial orders plotted as colored lines. The right panels show the best-fit baseline-subtracted spectrum, with the best-fit polynomial order written in the bottom left corner. The vertical gray dotted lines denote the line width of the disk, $\rm \Delta v_{disk}$, for the target galaxy, NGC\,891. The value of $\rm \Delta v_{candidate}$ (shown with vertical yellow bands) at the start and end of each iteration are noted as well. The fitting algorithm at this pointing converges in the fourth iteration. \textbf{3rd, 5th, and bottom row (from left to right):} $\rm \Delta v_{candidate}$, N(\hin), S/N, and number of velocity channels (=$\rm \Delta v_{candidate}/\delta v$) of the candidate emission for different $\rm \delta v$. The gray horizontal band denotes $\rm \Delta v_{disk}$ of NGC\,891 in the leftmost panel. The filled and unfilled circles mark the upper and lower boundary of the candidate emission, respectively. In the rightmost panel, the horizontal line is drawn at 5, the threshold to identify a positive fluctuation as a candidate emission. The vertical dashed line in each panel corresponds to the optimum $\rm \delta v$.}
    \label{fig:baselines}
\end{figure*}
We use an iterative approach to fit and subtract polynomial baselines from our final averaged spectra. The goal of the technique is to identify the signal-free regions of the spectrum and then to fit and subtract the lowest-order polynomial that yields a good fit to these signal-free regions. To accomplish this we need to identify the likely velocity range of emission and then fit and subtract the baseline.

The procedure, which is illustrated for one pointing in Figure\,\ref{fig:baselines}, proceeds as follows:

\begin{enumerate}[wide=0pt, itemsep=-0.1 em]
\item To identify the velocity range of the possible signal, we create boxcar smoothed versions of the spectrum at a series of progressively lower velocity resolutions from $\delta v = 5$ to $20$\,km\,s$^{-1}$ in steps of 2.5\,km\,s$^{-1}$. Smoothing reduces the noise in the spectrum and makes it easier to identify possible signals at the expense of losing information on the velocity structure of emission.

\item For each smoothed version of the spectrum, we start with a large velocity range $\rm -1000\leqslant v  < 1000\;km\;s^{-1}$
to estimate the baseline. We mask the emission from the Milky Way so that it does not contaminate the baseline. 



\item To determine an approximate velocity region of a potential \hi emission, $ \Delta \rm v_{{candidate}}$, we fit the whole baseline without excluding the velocity region of the \hi disk emission ($\rm \Delta v_{disk} = \pm$250 km s$^{-1}$ for NGC\,891 and $\pm$300 km s$^{-1}$ for NGC\,4565). We call this best fit a ``temporary baseline''. 

For the fitting, we use a robust polynomial fitting algorithm \texttt{imodpoly} from \texttt{\href{https://pybaselines.readthedocs.io/en/latest/introduction.html}{pybaseline}}.  
\texttt{modpoly} iteratively fits a polynomial baseline to the data using \textit{thresholding} to reject outliers. 
\textit{Thresholding} first fits the data using traditional least-squares and then sets the input data of the next iteration as the element-wise minimum between the current data and the current fit. \texttt{imodpoly} is an improved \texttt{modpoly} algorithm for noisy data that includes the standard deviation of the residual (= data -- baseline) when performing the \textit{thresholding}. 

We restrict our fits to use only low-order ($0\leqslant n\leqslant3$) polynomials to avoid over-fitting and under- or over-estimating the emission we are looking for. Then the best-fit polynomial order for the working $\delta \rm v$, $n_{\rm fit}$, is decided based on the smallest $\chi^2/dof$, supported by the qualitative judgment from inspecting the results by eye.  

\item In the ``temporary baseline''-subtracted spectrum, we identify the initial $\Delta \rm v_{{candidate}}$ by eye. 
We identify a positive fluctuation as a candidate emission when the velocity width of that fluctuation is at least 4--5 times the velocity resolution, e.g., a $>$20--25\,km s$^{-1}$ wide positive fluctuation is considered a candidate emission in a $\delta$v = 5\,km s$^{-1}$ spectrum. 

\item We return to the smoothed spectrum in step \# 2, and fit a new baseline to the spectrum. This time we exclude the $\Delta \rm v_{{candidate}}$ region from fitting if the candidate emission is not at rest (i.e., not centered around 0 km s$^{-1}$). If the candidate emission is at rest, the algorithm can identify it anyway. For example, we do not exclude any velocity region for all minor axis pointings of NGC\,891 (except UP3N, where the 50--150 km s$^{-1}$ region is excluded). For the major axis pointings of NGC\,891, we exclude the 0--250 km s$^{-1}$ region at lower declination (`DN'), and the -250 to 0 km s$^{-1}$ region at higher declination (`UP', except `UP3J', where we exclude the 100--400 km s$^{-1}$ region). 

If needed, we adjust the velocity range used for the baseline fit $ \Delta$v$_{\rm base}$ to ensure that it is larger than $\Delta \rm v_{candidate}$ and includes enough spectral range for a good fit. This best-fit baseline is the ``intermediate baseline''.

\item We calculate the velocity range over which $\rm T_B$ in the ``intermediate baseline''-subtracted spectrum is continuously positive around the center of the $\Delta \rm v_{candidate}$ region. That velocity range is set to be the new $\Delta \rm v_{candidate}$. Because we look for \hi emission bound to the halo, the maximum allowed candidate velocity range is set by $\rm \Delta v_{disk}$ (except `UP3J' of NGC\,891). 

\item With the new exclusion region defined in step \# 6, we repeat step \#\,5, this time using the ``intermediate baseline''-subtracted spectrum as input. We repeat steps \#5 and \#6 until $\Delta \rm v_{candidate}$ converge over successive iterations. Then the converged $\Delta \rm v_{candidate}$ is taken as the velocity range of detected \hi emission, $\Delta$v$_{\rm emit}$, for the working $\delta \rm v$. The best-fit baseline of the converged fit is the ``final baseline''. 

\item We calculate the RMS noise, $\sigma_{\rm T_B}$ from the signal-free part of the ``final baseline''-subtracted spectrum. Then, we integrate the spectrum over the range of $\Delta$v$_{\rm emit}$ to obtain the intensity, $\rm I = \int T_B dv$ in units of K km s$^{-1}$. We calculate the statistical uncertainty in the integrated intensity, $\rm \sigma_I = \sigma_{T_B}\sqrt{\delta v\Delta v_{emit}}$.

\item We repeat steps \#\,2 to 8 for different values of $\delta \rm v$.

\item We choose the optimum velocity resolution, the final $\delta$v, to be the value where the \hi emission is detected at maximum signal-to-noise ratio (S/N = I/$\sigma_{\rm I}$) in the smoothed and final-baseline-subtracted spectrum. 
\end{enumerate}

In Table \ref{tab:obs}, we provide the optimum velocity resolution, the final velocity range of \hi emission $\Delta$v$_{\rm emit}$, the final velocity range of baseline $\Delta$v$_{\rm base}$, and the best-fit polynomial orders $n_{\rm fit}$ that we find for each pointing.   

\begin{figure}
    \centering
    \includegraphics[width=0.48\textwidth]{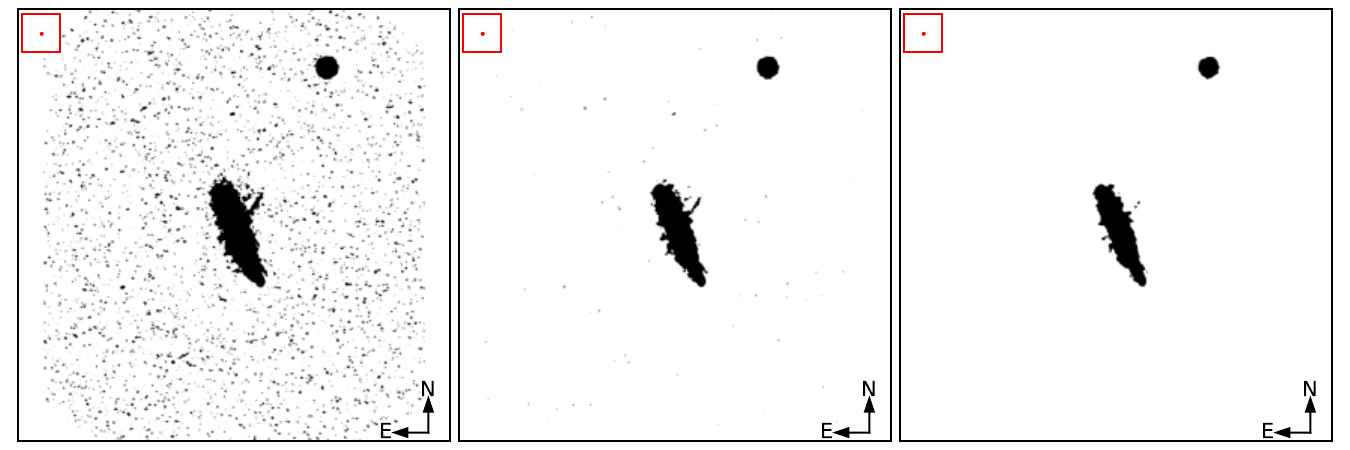}
    \includegraphics[width=0.48\textwidth]{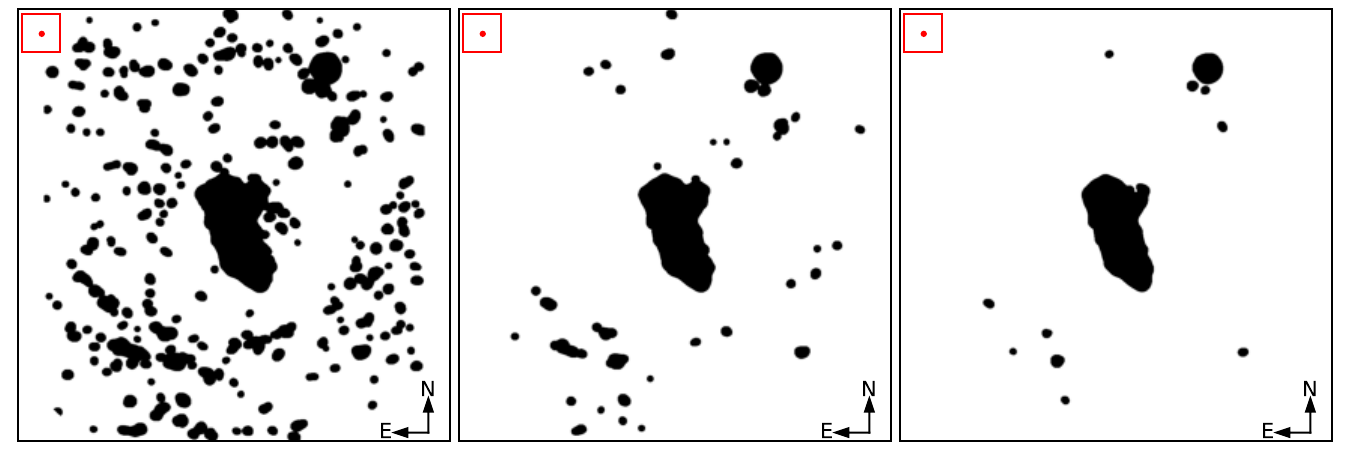}
    \includegraphics[width=0.48\textwidth]{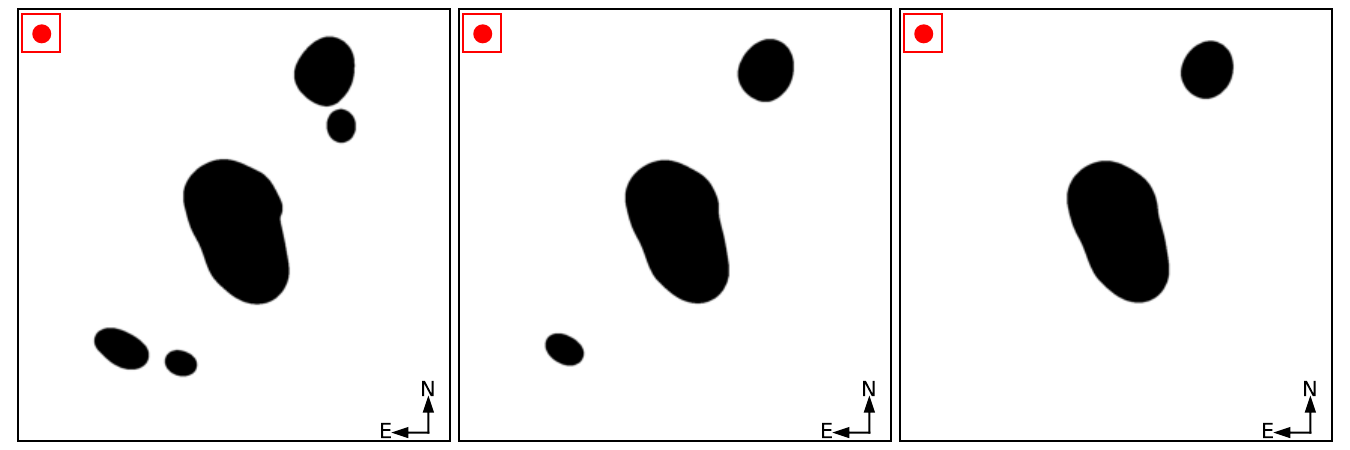}
    \includegraphics[width=0.48\textwidth]{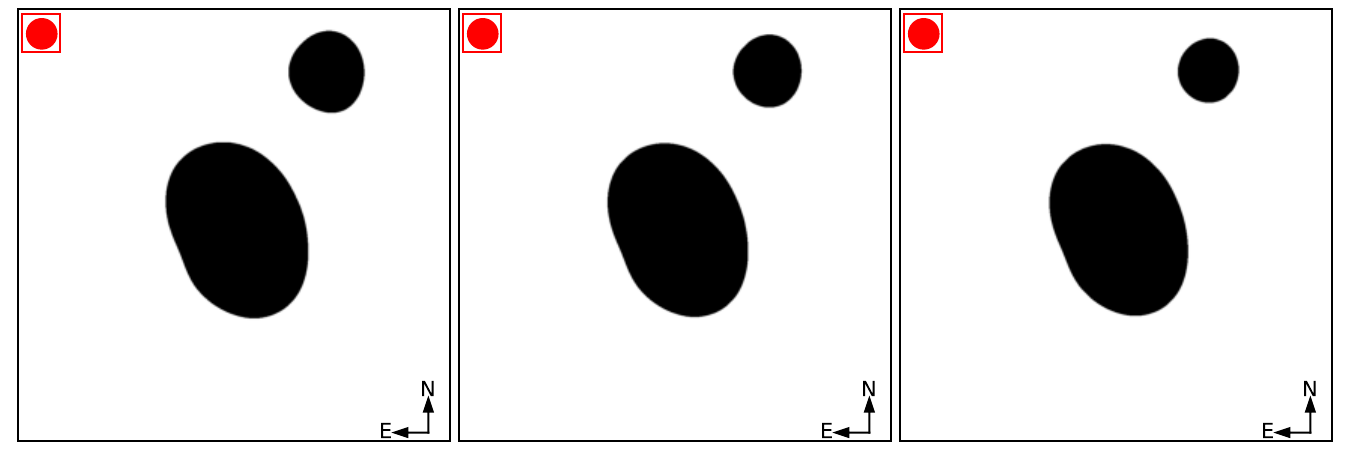}    
    \caption{Mask cubes summed over all velocity channels of the WSRT data cube of NGC\,891, for S/N$>$3 (left), S/N$>$4 (middle), and S/N$>$5 (right). Black regions with a value of 1 are extracted from the data cube. \textbf{Top row:} Mask for the original data cube, where pixels above the required S/N in two adjacent velocity channels are retained in the masked data cube. In addition to the \hi disk of NGC\,891, the filament, and UGC\,1807, there are a few pixels in the CGM with S/N$>$4. However, S/N$>$3 masking allows a lot of pixels across the field which is likely spurious. This was one of the reasons for choosing S/N$>4-5-6$ masking in Appendix\,\ref{sec:threshold} instead of S/N$>3-4-5$ as shown here. \textbf{Bottom three rows:} Mask based on the noise estimated from the data cube that is convolved with $1'$ (second row), $3'$ (third row), and $5'$ (fourth row) Gaussian beams. At every velocity channel, pixels above the required S/N and all pixels within $1'$, $3'$, or $5'$ of those pixels are retained in the masked data cube, respectively.}
    \label{fig:mask}
\end{figure}

\textcolor{black}{\section{Deciding Threshold for masking WSRT data cubes}}\label{sec:threshold}
\begin{figure}
    \centering
    \includegraphics[width=0.475\textwidth]{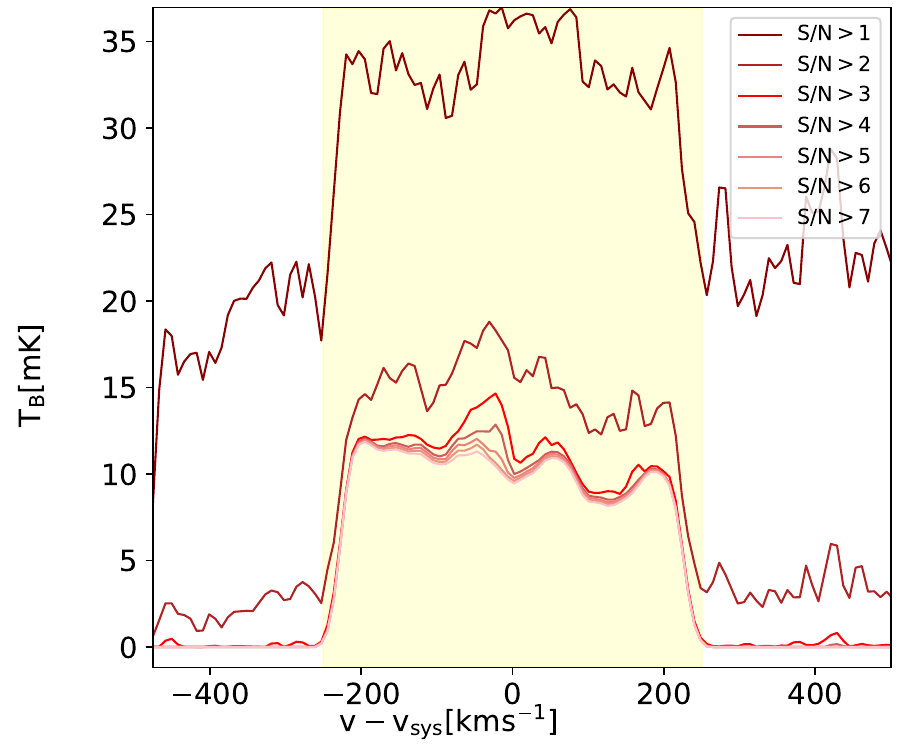}
    \vspace{-0.1 in}
    \caption{The WSRT spectrum toward the UP\,1N pointing of NGC\,891 convolved with the GBT beam. The legend details the S/N thresholds to mask the WSRT data cube before convolving. The opacity of the red shade is inversely proportional to the S/N threshold. The yellow vertical patch depicts the velocity range of the \hi disk.} 
    \label{fig:threshold}
\end{figure}
\begin{enumerate}[wide=0pt]
\item \textbf{Noise Calculation:}
To estimate the noise in the WSRT data cubes, we consider two ``emission-free” regions by excluding data within a radius of 20$'$ and 30$'$ from the center of the target galaxy. A 20$'$ mask is appropriate because signals can be visually identified within this radius, although some of our pointings lie beyond it. A 30$'$ mask provides a truer representation of the ``emission-free'' region, although it leaves a smaller amount of data to estimate the noise. Thus there is a trade-off between the accuracy and amount of the data used to estimate the noise. Because the positive values in the ``emission-free'' region might be affected by the faint (and visually unidentifiable) signals, we fit the histogram of only negative values in the ``emission-free'' region with a Gaussian. We consider the best-fitted standard deviation of that Gaussian as the global noise of the data cube. In the following steps, We consider the minimum of the noise values calculated using the two ``emission-free'' regions. 

\item \textbf{Masking:} 
Spectra from the unmasked and convolved WSRT cube would be heavily contaminated by noise, especially at the pointings more than 1 GBT beamsize away from the \hi disk. It might lead to the WSRT column densities being significantly larger than the GBT column densities, which is physically impossible. That makes it useless to compare unmasked and convolved WSRT spectra with the GBT spectra.

To mask the WSRT data cubes based on noise calculated in the previous step, we consider the S/N of each pixel in each velocity channel. We retain the value of a pixel at a given velocity channel if the S/N of that pixel at that velocity channel and in at least one of the adjacent velocity channels is above the required threshold. This allows us to avoid spurious emissions at a deserted velocity channel. We consider thresholds of 1--7 at a step of 1. Masked pixels in the cube are rendered with a value of zero. 

\item \textbf{Convolution and Spectra Extraction:}
We convolve the WSRT cubes masked at different thresholds with the circularized GBT beam and extract the spectrum at each pointing. In Fig.\,\ref{fig:threshold}, we show the convolved WSRT spectrum toward the UP\,1N pointing of NGC\,891 from the WSRT cubed masked at seven different masking thresholds. 

The spectra corresponding to the cube masked at S/N $\geqslant1$ and 2 are discarded since the baseline is far from zero and is systematically positive by construction. In the S/N $\geqslant3$ masking, the baseline estimate is significantly improved. But it still retains some residual noise in the cube (see the mask in Fig.\,\ref{fig:mask}, top row, left panel), leading to the residual emission in the convolved spectrum. While it is visible in the baseline region in contrast to the expected constant zero value of the baseline, it can affect any velocity channel, thus leading to an overestimated column density of the WSRT spectrum. It becomes more problematic for pointings farther out from the disk (Fig.\,\ref{fig:diffspec}). Therefore, we discard this threshold as well. Additionally, S/N $\geqslant7$ does not offer any significant improvement compared to S/N $\geqslant6$ and is therefore discarded. Thus, we discard thresholds of S/N$\geqslant$ 1--3 and $\geqslant7$, while retaining S/N  $\geqslant4-6$.

We consider the spectrum from the cube masked at S/N$\geqslant$5 as the best estimate. We use the spectra from S/N $\geqslant$ 4 and 6 masked cubes to calculate the additive systematic uncertainty. 


\end{enumerate}

\section{Difference spectra}
\begin{figure*}
    \centering
    \includegraphics[width=0.98\textwidth]{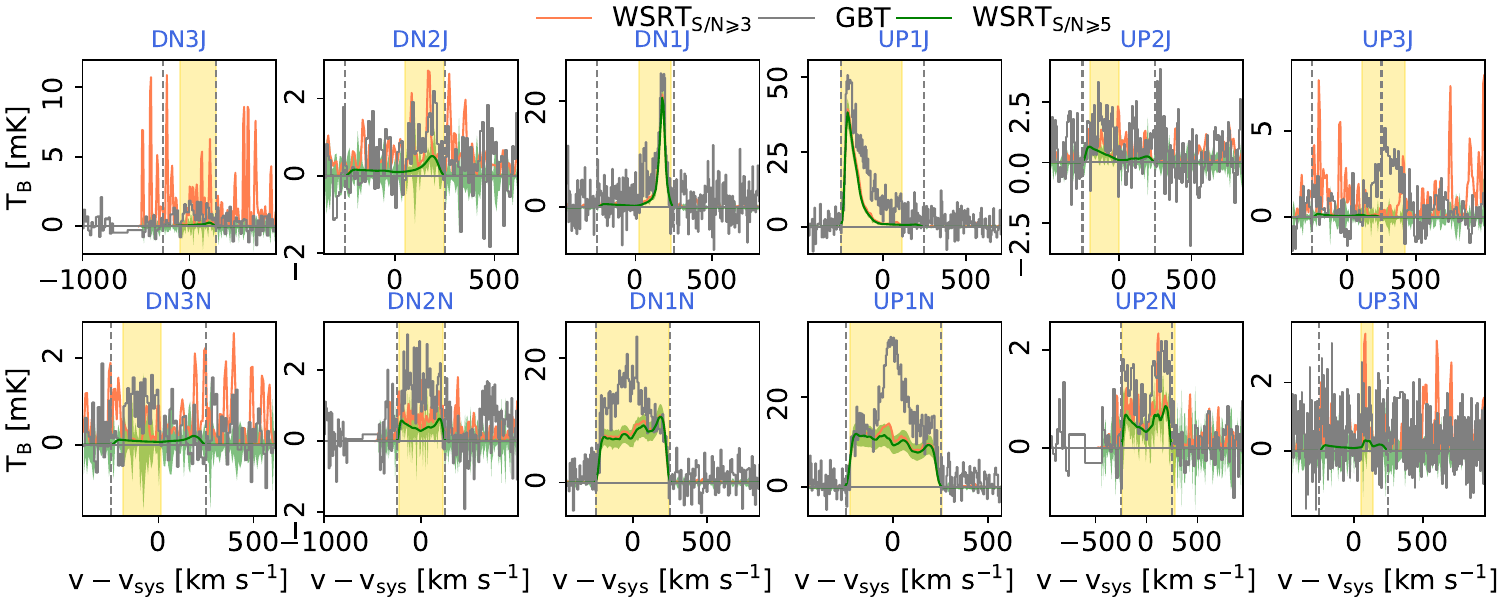}
    \caption{The GBT and WSRT spectra for our observed GBT pointings of NGC\,891. 
    The vertical yellow patch is the region where the \hi emission is detected in the GBT data. 
    The maximum rotational velocity of the \hi disk is marked with vertical dashed gray lines. 
    \textcolor{black}{The green and red curves correspond to WSRT cubes masked at S/N$>$5 and S/N$>$3, respectively, and convolved with the circularized GBT beam. The vertical extent of the green shaded regions around the green curves is due to masking the WSRT data cube at S/N$>4$ (upper extent) or S/N$>6$ (lower extent), and it includes 1) statistical uncertainty in WSRT data, and 2) systematic uncertainty due to the azimuthal asymmetry of the GBT beam}. The masking at S/N$>3$ (or any lower threshold) is discarded because it picks up a lot of spurious noise at UP/DN\,2 and 3 pointings, and does not pick up any true excess than the S/N$>4$ masking at UP/DN\,1 pointings.}
    \label{fig:diffspec}
\end{figure*}
The masked and convolved WSRT spectra in comparison with GBT spectra of NGC\,891 are shown in Figure\,\ref{fig:diffspec}. Qualitatively, the results look similar for NGC\,4565. 

\begin{figure}
    \centering
    \includegraphics[width= 0.48\textwidth]{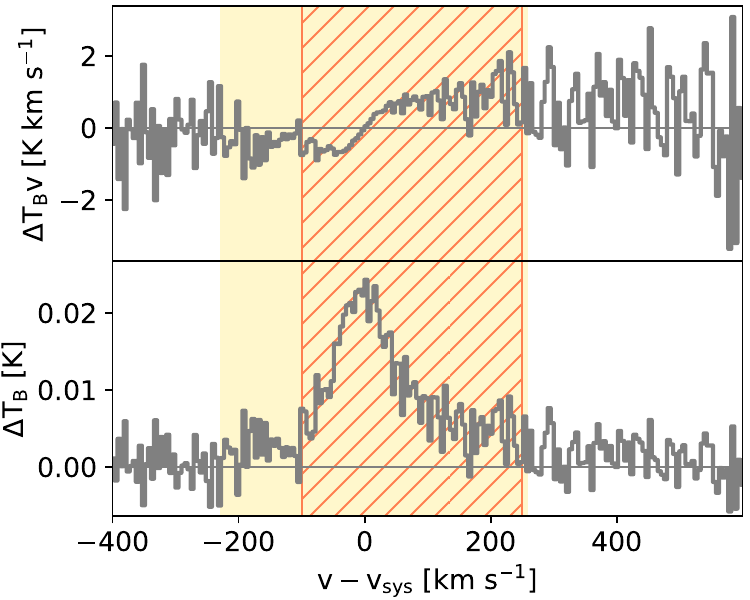}
    \caption{The difference in brightness temperature weighted velocity (top) and brightness temperature (bottom) between our GBT data and convolved WSRT data of the UP\,1N pointing of NGC\,891. The hatched region shows where the GBT has excess detection than WSRT. The area under the curve within the hatched region in the bottom panel is the excess integrated intensity of \hi detected by the GBT. The ratio of areas under the curves within the hatched regions in the top and bottom panels is the mean line-of-sight velocity of that excess \hin.}
    \label{fig:Tvdiffer}
\end{figure}
We show an example of the channel-wise difference between GBT and WSRT spectra for UP\,1N pointing of NGC\,891 in Figure\,\ref{fig:Tvdiffer}. Because the velocity channels of GBT and WSRT are not the same, we reconstruct the WSRT spectrum at the velocity channels of GBT using linear interpolation. In the top and bottom panels, we show the difference in brightness temperature weighted velocity and the brightness temperature.

\section{Intermediate Convolution}\label{sec:intermediate_convol}
Traditionally, masking involves comparing pixel values to an S/N threshold. However, we propose masking with intermediate convolution. This process is similar to traditional convolution but we introduce angular scale as an additional parameter. Instead of masking pixel values directly, we retain every value within a 3 arcmin radius. The radius of inclusion is determined as half the FWHM (beam size). To estimate the noise, we employ intermediate convolution with a Gaussian kernel on the data cube. 

\begin{figure}
    \centering
    \includegraphics[width=0.48\textwidth]{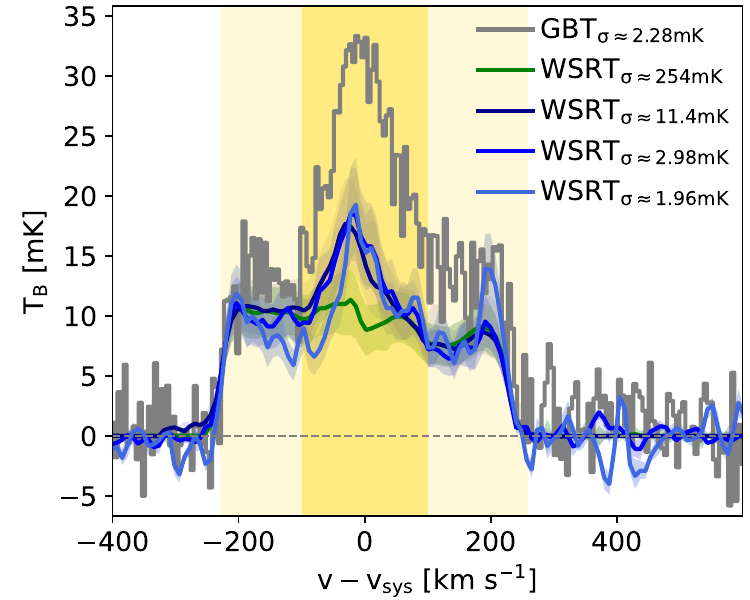}
    \caption{The spectra toward the UP\,1N pointing of NGC\,891 measured by the GBT and WSRT. The WSRT spectra are from convolved WSRT data cubes masked at S/N $>4$. The shaded regions include systematic uncertainties due to masking the cube at S/N $>3$ or S/N$>5$ and circularizing the azimuthally asymmetric GBT beam, and statistical uncertainty. The green curve is where the WSRT data cube has been directly masked and convolved (discussed in Appendix\,\ref{sec:threshold}). WSRT data cubes that are masked based on intermediate convolution with Gaussian beams are shown in different shades of blue, with dark to light shades denoting smaller to larger Gaussian beams. The statistical noise in the brightness temperature at a velocity resolution of 5 km\,s$^{-1}$, $\rm \sigma_{T_{B,5}}$, of each spectrum has been noted in the legend. }
    \label{fig:UP1N}
\end{figure}

We convolve the WSRT data cube with a Gaussian beam of FWHM = $r'$. We calculate the noise of that intermediate convolved cube in the same way as in the previous section. We mask the original cube by retaining pixels at every velocity channel with values above a certain S/N threshold depending on that noise and also all other pixels within $r'$ of those pixels. We try $r$ = 1, 3, and 5. The larger the $r$, the smaller the noise, and the higher the chance to extract more low-intensity genuine (not spurious) \hi emission from the WSRT data cube that could have been removed in the direct masking process discussed in the previous section. 

We convolve these masked-based-on-intermediate-convolution WSRT data cubes with the GBT beam and obtain the convolved spectra at our observed GBT pointings. At the same pointing, the difference in the spectra for different levels of intermediate convolution, if any, would be an indirect constraint on the angular scale of the detected \hi emission. We compare the shape and column density of these spectra with our GBT spectra and the directly masked and convolved WSRT spectra.

We show an example for the UP\,1N pointing in Figure\,\ref{fig:UP1N}. The shape of the GBT spectrum (gray curve) can be decomposed into two components: a \textit{broad emission} due to the \hi disk contamination (pale yellow patch), and a \textit{narrow emission} (bright yellow patch) possibly connected to the filament (visible in the green contours in Figure\,\ref{fig:spectra891}). All WSRT spectra can explain the \textit{broad emission}, as expected, but the \textit{narrow emission} is not captured by the directly masked and convolved WSRT spectrum (green curve in Figure\,\ref{fig:UP1N}). Because the WSRT data cube is two orders of magnitude shallower than our GBT observation (statistical noise level noted in the legend of Figure\,\ref{fig:UP1N}), we cannot rule out the possibility of the excess emission detected by GBT being small-scale low column density \hi that is hidden under the noise of the WSRT data cube but is missed through the masking process. The masking based on intermediate convolution allows us to extract some of that low column density \hi and thus retrieve a fraction of the \textit{narrow emission} (blue curves in Figure\,\ref{fig:UP1N}). The N(\hin) of these WSRT spectra are $8.8^{+2.3}_{-1.8} \times 10^{18}$, $8.7^{+1.9}_{-1.8} \times 10^{18}$, and $8.4\pm 1.8 \times 10^{18}$ cm$^{-2}$ (for intermediate convolution with $1', 3'$ and $5'$ Gaussian beams, respectively). These are consistent with the N(\hin) of the directly masked and convolved WSRT spectrum, $7.9_{-1.8}^{+1.7} \times 10^{18}$ cm$^{-2}$, within $1\sigma$ uncertainty, thus making a negligible difference despite the difference in shape. The noise level of the WSRT data cube convolved with $5'$ Gaussian beam is lower than that of our GBT spectrum, yet we can see that a large fraction of the \textit{narrow emission} remains unexplained. This is a rigorous confirmation of the excess emission detected by our GBT observation being a diffuse extended \hi and not small-scale clumps/clouds.

\bsp	
\label{lastpage}
\end{document}